\documentclass[journal, onecolumn, 12pt, draftclsnofoot,romanappendices]{IEEEtran}
\usepackage{cite}
\usepackage{graphicx,color,overpic,psfrag}
\usepackage{subfigure}
\usepackage{amsmath}
\usepackage{times}
\usepackage{latexsym}
\usepackage{bm}
\usepackage{amssymb}
\usepackage{cases}
\usepackage{array}
\usepackage{fancyhdr}
\usepackage{stfloats}
\usepackage{setspace}
\usepackage{multirow}
\usepackage{slashbox}
\usepackage{psfrag}
\usepackage{color}
\usepackage{caption}
\usepackage{algorithm}
\usepackage{algorithmicx}
\usepackage{algpseudocode}

\newtheorem{theorem}{Theorem}

\newtheorem{lemma}{Lemma}

\begin{document}
%
\title{Uplink Spectral Efficiency Analysis of Multi-Cell Multi-User Massive MIMO over Correlated Ricean Channel}
%
%
%

\author{Juan Cao,
        Dongming Wang,
        Jiamin Li
        and Qiang Sun}
\maketitle

\begin{abstract}
In this paper, the performance of uplink spectral efficiency in massive multiple input multiple output (MIMO) over spacially correlated Ricean fading channel is presented. The maximum ratio combining (MRC) receiver is employed at the base station (BS) for two different methods of channel estimation. The first method is based on pilot-assisted least minimum mean square error (LMMSE) estimation, the second one is based on line-of-sight (LOS) part. The respective closed-form expressions of uplink data rate are given for these two methods. Due to the existence of pilot contamination, the uplink data rate of pilot-assisted LMMSE estimation method approaches to a finite value (we name it as asymptotic rate in the paper) when the BS antenna number is very large. However, the data rate of LOS method goes linearly with the number of BS antennas. The expression of the uplink rate of LOS method also show that for Ricean channel, the spacial correlation between the BS antennas may not only decrease the rate, but also increase the rate, which depends on the locations of the users. This conclusion explains why the spacial correlation may increase, rather than decrease the data rate of pilot-assisted LMMSE. We also discuss the power scaling law of the two methods, and the asymptotic expressions of the two methods are the same and both independent of the antenna correlation.
\end{abstract}

\begin{IEEEkeywords}
Massive {massive {MIMO}, pilot contamination, Ricean fading, spacial correlation, uplink rate}
\end{IEEEkeywords}

%
\IEEEpeerreviewmaketitle

\section{Introduction}
Massive multiple input multiple output (MIMO) technology has been investigated with the remarkable potential of increasing the spectral efficiency and energy efficiency even with very simple linear transmitter/receiver \cite{lmimot,lsfw,ica,vtc17}. However, increasing antenna number leads to new challenges such as how to obtain the accurate channel state information (CSI). For time division duplexing (TDD) systems, the downlink CSI can be obtained by using uplink pilots and exploiting channel reciprocity. However, there will be estimation errors, feedback delay errors and quantized errors which impair the system performance. Besides, limited to the coherence time, the number of the orthogonal pilots is limited. And we have to reuse the same pilot sequences in different cells  which causes pilot contamination. The system capacity of massive-MIMO is hardly decided by the reuse of pilot sequences between the nearby cells \cite{lmimothy,Ltepilot}. How to reduce the pilot contamination is crucial to further improve massive-MIMO performances.

\cite{Josepilot} proposed a new multi-cell minimum mean square error (MMSE) based precoding method for the sake of reducing pilot contamination. For TDD system, the overall cells are divided into two groups, the users in the cells of the same group transmit pilot sequence to the base stations (BSes) while users in cells of other group receive data. By this way pilot contamination only remains within the same group and is efficiently reduced without inter-cell coorparation\cite{Pilotyi,pilotFab}. In such technologies, the SINR of user is limited and can not go to infinity with the increase of the number of BS antennas. On the other hand, the effects of the pilot contamination is related to the pilot assignment scheme. Hereby, many researchers studied how to optimize the pilot assignment or pilot scheduling schemes\cite{schsan,schwang,schtwo,schesi}. Besides, some works indicated that pilot contamination isn't inevitable as we expected. For example, in the case covariance matrices satisfy a certain non-overlapping condition on their dominant subspaces, a Bayesian channel estimation method making explicit use of covariance information can completely remove pilot contamination effects\cite{Pilotyin}. Since the source of pilot contamination lies in using pilot assisted channel estimation,  the channel can be estimated blindly. Herein, the pilot contamination is avoided as no pilots are used
\cite{Bilndngo,Blindmuller,Blindmullert}. Taking the inherent sparsity of wireless channel into consideration, sparsity channel estimation and pilot design can be employed using  compressive sensing technology to reduce pilot overhead and pilot contamination\cite{Comprdavid,Comprqi,Compreqit}.

The other challenge of massive MIMO lies in the limited space at BS-side which causes antenna mounting to become a difficult issue. And millimeter-wave (mm-wave) wireless systems are emerging as a promising technology for full exploitation of spacial multiplex and development of higher spectrum. The millimeter-wave operates from 30 to 300 GHz with wavelength between 10mm to 1mm, and the smaller wavelength makes it possible that large number of antennas can be mounted in the limited space\cite{lossay,losbra}. Due to the highly directional nature of propagation, line-of-sight (LOS) propagation plays an important role at mm-wave. Besides, when the cell coverage is shrunk, the channel between users and the BS will highly probable to include LOS part \cite{smacel,femcel}. As a result, the LOS is expected to be a new propagation mode for the massive MIMO. However, the majority of works are based on the assumption of Rayleigh fading model which simplify the mathematic model and analysis. The Rayleigh model will no longer suit when there's LOS. The Ricean fading model is  applicable when the wireless link between the transmitter and the receiver has LOS component in addition to the diffused Rayleigh component. \cite{cmrice} investigated the uplink rate of massive MIMO over i.i.d. Ricean fading channels.

In this paper, we investigate the uplink data rate of massive MIMO over correlated Ricean fading channels for two CSI estimation methods. The first method is maximum ratio combining (MRC) based on pilot-assisted LMMSE estimation which will cause pilot contamination. In order to avoid the pilot contamination, similar to \cite{tjcsi}, we take the first order statistical information as the estimated channel, which means taking the LOS part as the estimated channel while the diffused part is regarded as interference. The main contributions are:

(1) We deduce the respective analytical expressions for two methods. Due to the existence of pilot contamination,
the uplink data rate of pilot-assisted LMMSE estimation method approaches to a finite value when the BS antenna number is very large. However, the infinite uplink achievable rate of LOS method goes linearly with the number of BS antennas, which means that as the number of BS antennas increase, the gap between the two methods will become smaller and smaller, and finally the rate of LOS method will exceed the one of pilot-assisted LMMSE estimation method.

(2) The impact of Ricean fading has been investigated extensively. According to the expression of the achievable rate of LOS method, we find that the correlation between the BS antennas may not only decrease the rate, but also increase the rate, which depends on the locations of the users. So if the locations of the users make antenna correlation increase the rate, with the increase of Ricean factor the rate of pilot-assisted LMMSE estimation method will become larger due to antenna correlation since the effect of LOS part becomes more and more strong.

(3) We also discuss the power scaling law of the two methods. When the base station antenna number is very large, the asymptotical expressions of the two methods are the same and independent of the antenna correlation.


\section{System model}
\label{sec:1}
Consider the $L$ cells system with one BS and $K$ mobile users in each cell. Each BS is equipped with $N$ antennas, and each user has a single antenna. We assume that the system is operating on TDD protocol with full frequency reuse. Taking cell $1$ as the reference cell, the uplink received base-band signal vector is given by

\begin{equation}\label{eq:receive_model}
{\bm{y}} = \sqrt {{p_u}} {{\bm{G}}_1}{{\bm{x}}_1} + \sqrt {{p_u}} \sum\limits_{l = 2}^L {{{\bm{G}}_l}{{\bm{x}}_l}}  + {\bm{w}},
\end{equation}
where ${\bm{y}} = {\left[ {{y_1} \cdots {y_N}} \right]^{\rm{T}}}$ is the received signal vector, ${{\bm{x}}_l} = {\left[ {{x_{l,1}} \cdots {x_{l,K}}} \right]^{\rm{T}}} \sim {\cal C}{\cal N}\left( {{\bm{0}},{{\bm{I}}_K}} \right)$, where ${x_{l,k}}$ is the transmitted signal of user $k$ in $l$th cell, and $p_u$ is the average transmit power of each user. ${{\bm{G}}_l} = \left[ {{{\bm{g}}_{l,1}} \cdots {{\bm{g}}_{l,K}}} \right]$ is the composite channel matrix of all the $K$ users in $l$th cell to the reference cell. ${\bm{w}}$ is the additive noise vector which satisfies standard complex Gaussian distribution. Since the distance between the users in the reference cell and the reference BS is small, the channel of users in the reference cell is modeled to consist of two parts, namely a deterministic component corresponding to the LOS path and a Rayleigh-distributed random component which accounts for the scattered signals. By contrast, the distance between users in the interfering cell and the reference BS is large, so the LOS part does not exist any more in channels because of the scatters and buildings block. Based on this, the fast fading can be modelled as
\begin{equation}
{{\bm{H}}_l} =
 \left\{
 \begin{array}{ll}
{{{\bm{\bar H}}}_1}{\left[ {{\bm{\Omega }}{{\left( {{\bm{\Omega }} + {{\bm{I}}_K}} \right)}^{ - 1}}} \right]^{\frac{1}{2}}} + {\bm{\mathord{\buildrel{\lower3pt\hbox{$\scriptscriptstyle\frown$}}\over H} }}_1{\left[ {{{\left( {{\bm{\Omega }} + {{\bm{I}}_K}} \right)}^{ - 1}}} \right]^{\frac{1}{2}}}&l = 1\\
{\bm{\mathord{\buildrel{\lower3pt\hbox{$\scriptscriptstyle\frown$}}\over H} }_l}&{\kern 1pt} l \ne 1
\end{array}
\right.,
\end{equation}
where ${\left[ {{{{\bm{\bar H}}}_1}} \right]_{N \times K}}$ is the deterministic component with ${\left[ {{{{\bm{\bar H}}}_1}} \right]_{n,k}} = {e^{ - j\left( {n - 1} \right)\frac{{2\pi d}}{\lambda }\sin {\theta _k}}}$, and $d$ is the antenna spacing, $\lambda$ is the wavelength, ${\theta _k} \sim \left[ { - {\pi  \mathord{\left/
 {\vphantom {\pi  2}} \right.
 \kern-\nulldelimiterspace} 2},{\pi  \mathord{\left/
 {\vphantom {\pi  2}} \right.
 \kern-\nulldelimiterspace} 2}} \right]$ is the arrival angle of $k$th user in the reference cell. ${\left[ {{{{\bm{\mathord{\buildrel{\lower3pt\hbox{$\scriptscriptstyle\frown$}}
\over H} }}}_l}} \right]_{N \times K}} = \left[ {{{\bm{h}}_{l,1}}{{\bm{h}}_{l,2}} \cdots {{\bm{h}}_{l,K}}} \right]$ denotes the channel matrix for the fast fading between the users in each cell and the reference BS which satisfies the standard complex Gaussian distribution. ${\bm{\Omega }}$ is a diagonal matrix with ${{\bm{\Omega }}_{k,k}} = {\vartheta _k}$ as the $k$th element denoting the Ricean factor representing the ratio of the power of the deterministic component to the power of the fading component. And the bigger ${\vartheta _k}$ is, the more deterministic the channel is. Taking the correlation between the BS antennas into consideration and assuming all users'  correlation are the same, the composite channel matrix can be expressed as

\begin{equation}
{{\bm{G}}_l} = \left\{ \begin{array}{ll}
{{{\bm{\bar G}}}_1}{\left[ {{\bm{\Omega }}{{\left( {{\bm{\Omega }} + {{\bm{I}}_K}} \right)}^{ - 1}}} \right]^{\frac{1}{2}}} + {{{\bm{\mathord{\buildrel{\lower3pt\hbox{$\scriptscriptstyle\frown$}}
\over G} }}}_1}{\left[ {{{\left( {{\bm{\Omega }} + {{\bm{I}}_K}} \right)}^{ - 1}}} \right]^{\frac{1}{2}}} & l = 1\\
{{{\bm{\mathord{\buildrel{\lower3pt\hbox{$\scriptscriptstyle\frown$}}
\over G} }}}_l} & l \ne 1
\end{array} \right.,
\end{equation}
where
\[{{\bm{\bar G}}_1} = \left[ {{{{\bm{\bar g}}}_{1,1}} \cdots {{{\bm{\bar g}}}_{1,K}}} \right] = {{\bm{\bar H}}_1}{\bm{\Lambda }}_1^{\frac{1}{2}},\]
\[{{\bm{\mathord{\buildrel{\lower3pt\hbox{$\scriptscriptstyle\frown$}}
\over G} }}_l} = \left[ {{{{\bm{\mathord{\buildrel{\lower3pt\hbox{$\scriptscriptstyle\frown$}}
\over g} }}}_{l,1}}\cdots {{{\bm{\mathord{\buildrel{\lower3pt\hbox{$\scriptscriptstyle\frown$}}
\over g} }}}_{l,K}}} \right] = {{\bm{R}}^{\frac{1}{2}}}{{\bm{\mathord{\buildrel{\lower3pt\hbox{$\scriptscriptstyle\frown$}}
\over H} }}_l}{\bm{\Lambda }}_l^{\frac{1}{2}},\]
 ${{\bm{\Lambda }}_l} \buildrel \Delta \over = \text{diag} {\left( {\begin{array}{*{20}{c}}
{{\lambda _{l,1}}}& \cdots &{{\lambda _{l,K}}}
\end{array}} \right)} $ with ${\lambda _{l,k}}$ represents the slow fading(including the shadow and path loss), $\bm{R}$ is the deterministic receive correlation matrix, and $\bm{R}$ has properties as follow: positive definite; $\text{Tr}\left[ {\bm{R}} \right] = N$; having uniformly bounded spectral norm.

\section{Uplink rate analysis using pilot assisted LMMSE channel estimate}
\subsection{LMMSE channel estimate}
Assuming both the deterministic LOS component and the Ricean factor matrix ${\bm{\Omega }}$ are perfectly known at both the transmitter and the receiver, only the Rayleigh fading part needs to be estimated. Define the estimated channel matrix as:
\begin{equation}
{{\bm{\hat G}}_l} =
\left\{
 \begin{array}{ll}
{{{\bm{\bar G}}}_1}{\left[ {{\bm{\Omega }}{{\left( {{\bm{\Omega }} + {{\bm{I}}_K}} \right)}^{ - 1}}} \right]^{\frac{1}{2}}} + {{{\bm{\hat{\mathord{\buildrel{\lower1pt\hbox{$\scriptscriptstyle\frown$}}\over G}} }}}_1}{\left[ {{{\left( {{\bm{\Omega }} + {{\bm{I}}_K}} \right)}^{ - 1}}} \right]^{\frac{1}{2}}}& l = 1\\
{{{\bm{\hat {\mathord{\buildrel{\lower1pt\hbox{$\scriptscriptstyle\frown$}}\over G}} }}}_l} & l\ne 1
\end{array}
 \right..
 \end{equation}

In the multi-cell scenario, non-orthogonal training sequences must be used due to the limitation of orthogonal pilot resources which are limited by the coherence time of the channel. During the uplink pilot transmission, users in all cells simultaneously transmit pilot sequence of length $\tau $, and $K \le \tau  < T$, where $T$ is the coherence time of the channel. In this paper we let $\tau  = K$, so the training matrix is a $K \times K$ unitary matrix  satisfying ${{\bm{\Phi }}^{\rm{H}}}{\bm{\Phi }} = {{\bm{I}}_K}$.
The received pilot signals at reference cell can be expressed as
\[{{\bm{Y}}_{\rm{P}}} = \sqrt {{p_{\rm{P}}}} \left\{ {{{{\bm{\bar G}}}_1}{{\left[ {{\bm{\Omega }}{{\left( {{\bm{\Omega }} + {{\bm{I}}_K}} \right)}^{ - 1}}} \right]}^{\frac{1}{2}}} + {{{\bm{\mathord{\buildrel{\lower3pt\hbox{$\scriptscriptstyle\frown$}}
\over G} }}}_1}{{\left[ {{{\left( {{\bm{\Omega }} + {{\bm{I}}_K}} \right)}^{ - 1}}} \right]}^{\frac{1}{2}}}} \right\}{{\bm{\Phi }}^{\rm{T}}} + \sqrt {{p_{\rm{P}}}} \sum\limits_{l = 2}^L {{{{\bm{\mathord{\buildrel{\lower3pt\hbox{$\scriptscriptstyle\frown$}}
\over G} }}}_l}{{\bm{\Phi }}^{\rm{T}}}}  + {{\bm{W}}_{\rm{P}}},\]
where ${{\bm{W}}_{\rm{P}}}$  is a $N \times K$ noise matrix satisfying standard complex Gaussian distribution. Taking the LOS part off, we get
\[{\bm{Y}}_{\rm{P}}^* = \sqrt {{p_{\rm{P}}}} {{\bm{\mathord{\buildrel{\lower3pt\hbox{$\scriptscriptstyle\frown$}}
\over G} }}_1}{\left[ {{{\left( {{\bm{\Omega }} + {{\bm{I}}_K}} \right)}^{ - 1}}} \right]^{\frac{1}{2}}}{{\bm{\Phi }}^{\rm{T}}} + \sqrt {{p_{\rm{P}}}} \sum\limits_{l = 2}^L {{{{\bm{\mathord{\buildrel{\lower3pt\hbox{$\scriptscriptstyle\frown$}}
\over G} }}}_l}{{\bm{\Phi }}^{\rm{T}}}}  + {{\bm{W}}_{\rm{P}}}.\]
After correlating the received training signal ${\bm{Y}}_{\rm{P}}^*$ with the pilot sequence of user $k$, we get
 \[{{\bm{y}}_{{\rm{P}},k}^*} = \frac{{\sqrt {{p_{\rm{P}}}} }}{{\sqrt {{\vartheta _k} + 1} }}{{\bm{\mathord{\buildrel{\lower3pt\hbox{$\scriptscriptstyle\frown$}}
\over g} }}_{1,k}} + \sqrt {{p_{\rm{P}}}} \sum\limits_{l = 2}^L {{{{\bm{\mathord{\buildrel{\lower3pt\hbox{$\scriptscriptstyle\frown$}}
\over g} }}}_{l,k}}}  + {{\bm{w}}_{{\rm{P}},k}},\]
where ${{\bm{w}}_{{\rm{P}},k}} \sim {\cal C}{\cal N}\left( {{\bm{0}},{{\bm{I}}_N}} \right)$ is the addictive Gaussian white noise. So according to the LMMSE theory, we get
\[{{\bm{\hat {\mathord{\buildrel{\lower1pt\hbox{$\scriptscriptstyle\frown$}}
\over g} }}}_{l,k}}\,\, = \left\{ \begin{array}{ll}
{\frac{{{\lambda _{1,k}}{\bm{R}}}}{{\sqrt {{\vartheta _k} + 1} }}\frac{{{{\bm{Q}}_k}}}{{\sqrt {{p_{\rm{P}}}} }}{{\bm{y}}_{{\rm{P}},k}^*}} & l = 1\\
{{\lambda _{l,k}}{\bm{R}}\frac{{{{\bm{Q}}_k}}}{{\sqrt {{p_{\rm{P}}}} }}{{\bm{y}}_{{\rm{P}},k}^*}} &l \ne 1
\end{array} \right.,\]
where ${{\bm{Q}}_k} = {\left( {\frac{{{\lambda _{1,k}}{\bm{R}}}}{{{\vartheta _k} + 1}} + \sum\limits_{l = 2}^L {{\lambda _{l,k}}{\bm{R}}}  + \frac{{{{\bm{I}}_N}}}{{{p_P}}}} \right)^{ - 1}}$.
We define ${{\bm{\hat h}}_k} \buildrel \Delta \over = {\left( {\frac{{{{\bm{Q}}_k}}}{{{p_P}}}} \right)^{\frac{1}{2}}}{{\bm{y}}_{{\rm{P}},k}^*} \sim {\cal C}{\cal N}\left( {{\bm{0}},{{\bm{I}}_N}} \right)$ as the Rayleigh fading part of the estimated channel. Thus
${{\bm{\hat {\mathord{\buildrel{\lower1pt\hbox{$\scriptscriptstyle\frown$}}\over g} }}}_{l,k}}$ can be modelled as
\[{{\bm{\hat {\mathord{\buildrel{\lower1pt\hbox{$\scriptscriptstyle\frown$}}
\over g} }}}_{l,k}} = \left\{ \begin{array}{ll}
{\frac{{{\lambda _{1,k}}{\bm{R}}}}{{\sqrt {{\vartheta _k} + 1} }}{{\bm{Q}}_k}^{\frac{1}{2}}{{{\bm{\hat h}}}_k}}& l = 1\\
{{\lambda _{l,k}}{\bm{R}}{{\bm{Q}}_k}^{\frac{1}{2}}{{{\bm{\hat h}}}_k}} & l \ne 1
\end{array} \right..\]

Based on the orthogonality property of LMMSE estimate, ${{\bm{\mathord{\buildrel{\lower3pt\hbox{$\scriptscriptstyle\frown$}}
\over g} }}_{l,k}} $ can be decomposed as ${{\bm{\mathord{\buildrel{\lower3pt\hbox{$\scriptscriptstyle\frown$}}
\over g} }}_{l,k}} = {{\bm{\hat {\mathord{\buildrel{\lower1pt\hbox{$\scriptscriptstyle\frown$}}
\over g} }}}_{l,k}} + {{\bm{\tilde {\mathord{\buildrel{\lower1pt\hbox{$\scriptscriptstyle\frown$}}
\over g} }}}_{l,k}}$, where ${{\bm{\tilde {\mathord{\buildrel{\lower1pt\hbox{$\scriptscriptstyle\frown$}}
\over g} }}}_{l,k}}$ is the uncorrelated estimation error.
Since ${{\bm{\hat g}}_{1,k}} = \frac{{{{{\bm{\hat {\mathord{\buildrel{\lower1pt\hbox{$\scriptscriptstyle\frown$}}
\over g} }}}}_{1,k}}}}{{\sqrt {{\vartheta _k} + 1} }} + \frac{{\sqrt {{\vartheta _k}} }}{{\sqrt {{\vartheta _k} + 1} }}{{\bm{\bar g}}_{1,k}}$, ${{\bm{\tilde g}}_{1,k}} = \frac{{{{{\bm{\tilde {\mathord{\buildrel{\lower1pt\hbox{$\scriptscriptstyle\frown$}}
\over g} }}}}_{1,k}}}}{{\sqrt {{\vartheta _k} + 1} }}$, and ${{\bm{\hat g}}_{l,k}} = {{\bm{\hat{\mathord{\buildrel{\lower1pt\hbox{$\scriptscriptstyle\frown$}}\over g} }}}}_{l,k}$, ${{\bm{\tilde g}}_{l,k}} ={\bm{\tilde {\mathord{\buildrel{\lower1pt\hbox{$\scriptscriptstyle\frown$}}\over g} }}_{l,k}}$ when $l \ne 1$, the overall estimated channel can be expressed as
\begin{align}
{{\bm{\hat g}}_{l,k}} = \left\{
\begin{array}{ll}
\frac{{{\lambda _{1,k}}{\bm{R}}}}{{{\vartheta _k} + 1}}{{\bm{Q}}_k}^{\frac{1}{2}}{{{\bm{\hat h}}}_k} + \frac{{\sqrt {{\vartheta _k}} }}{{\sqrt {{\vartheta _k} + 1} }}{{{\bm{\bar g}}}_{1,k}}\quad & l= 1\\
{\lambda _{l,k}}{\bm{R}}{{\bm{Q}}_k}^{\frac{1}{2}}{{{\bm{\hat h}}}_k}&l \ne 1
\end{array} \right.,
\end{align}
and the covariance matrix of the overall estimation error is
\begin{align}
{\mathop{\rm cov}} \left( {{{{\bm{\tilde g}}}_{l,k}},{{{\bm{\tilde g}}}_{l,k}}} \right) = \left\{ \begin{array}{ll}
{\frac{{{\lambda _{1,k}}{\bm{R}}}}{{{\vartheta _k} + 1}} - {\left( {\frac{{{\lambda _{1,k}}}}{{{\vartheta _k} + 1}}} \right)^2}{\bm{R}}{{\bm{Q}}_k}{\bm{R}}} & {l=1}\\
{{\lambda _{l,k}}{\bm{R}} - \lambda _{l,k}^2{\bm{R}}{{\bm{Q}}_k}{\bm{R}}}& {l \ne 1}
\end{array} \right..
\end{align}

\subsection{Achievable uplink rate}
Considering the uplink transmission represented by \eqref{eq:receive_model}, the overall uplink received signal at the reference BS can be written as
\[{\bm{y}} = \sqrt {{p_u}} \sum\limits_{i = 1}^K {{{{\bm{\hat g}}}_{1,i}}} {x_{1,i}} + \sqrt {{p_u}} \sum\limits_{i = 1}^K {{{{\bm{\tilde g}}}_{1,i}}} {x_{1,i}} + \sqrt {{p_u}} \sum\limits_{l = 2}^L {\sum\limits_{i = 1}^K {{{{\bm{\hat g}}}_{l,i}}} {x_{l,i}}}  + \sqrt {{p_u}} \sum\limits_{l = 2}^L {\sum\limits_{i = 1}^K {{{{\bm{\tilde g}}}_{l,i}}{x_{l,i}}} }  + {\bm{w}}.\]
We use the linear receiver matrix which depends on the estimated channel for data detection. For MRC receiver,
\[{{\bm{c}}_k} = {\bm{\hat g}}_{1,k}^{\rm{H}} = {\left( {\frac{{{{{\bm{\hat {\mathord{\buildrel{\lower3pt\hbox{$\scriptscriptstyle\frown$}}
\over g} }}}}_{1,k}}}}{{\sqrt {{\vartheta _k} + 1} }} + \frac{{\sqrt {{\vartheta _k}} }}{{\sqrt {{\vartheta _k} + 1} }}{{{\bm{\bar g}}}_{1,k}}} \right)^{\rm{H}}}.\]
Thus we get
\[\begin{array}{l}
{r_k} = \sqrt {{p_u}} {\bm{\hat g}}_{1,k}^{\rm{H}}{{{\bm{\hat g}}}_{1,k}}{x_{1,k}} + \sqrt {{p_u}} {\bm{\hat g}}_{1,k}^{\rm{H}}\sum\limits_{i \ne k}^K {{{{\bm{\hat g}}}_{1,i}}} {x_{1,i}} + \sqrt {{p_u}} {\bm{\hat g}}_{1,k}^{\rm{H}}\sum\limits_{i = 1}^K {{{{\bm{\tilde g}}}_{1,i}}} {x_{1,i}}\\
\quad \quad \quad \quad  + \sqrt {{p_u}} \sum\limits_{l = 2}^L {\sum\limits_{i = 1}^K {{\bm{\hat g}}_{1,k}^{\rm{H}}{{{\bm{\hat g}}}_{l,i}}} {x_{l,i}}}  + \sqrt {{p_u}} \sum\limits_{l = 2}^L {\sum\limits_{i = 1}^K {{\bm{\hat g}}_{1,k}^{\rm{H}}{{{\bm{\tilde g}}}_{l,i}}{x_{l,i}}} }  + {\bm{\hat g}}_{1,k}^{\rm{H}}{\bm{w}}
\end{array}.\]
Then the achievable rate of user $k$ is
\[{R_k} = E\left[ {{{\log }_2}\left( {1 + \frac{{{{\left| {{\bm{\hat g}}_{1,k}^{\rm{H}}{{{\bm{\hat g}}}_{1,k}}} \right|}^2}}}{{{\bm{\hat g}}_{1,k}^{\rm{H}}\left( {\sum\limits_{l = 1}^L {\sum\limits_{i \ne k}^K {{{{\bm{\hat g}}}_{l,i}}} {\bm{\hat g}}_{l,i}^{\rm{H}}}  + \sum\limits_{l = 1}^L {\sum\limits_{i = 1}^K {{{{\bm{\tilde g}}}_{l,i}}{\bm{\tilde g}}_{l,i}^{\rm{H}}} }  + \frac{{{{\bm{I}}_N}}}{{{p_u}}} + \sum\limits_{l = 2}^L {{{{\bm{\hat g}}}_{l,k}}{\bm{\hat g}}_{l,k}^{\rm{H}}} } \right){{{\bm{\hat g}}}_{1,k}}}}} \right)} \right].\]

According to Lemma 1, the achievable rate of user $k$ can be approximated simply as follow.

\begin{theorem}\label{thm:lmmse}
For system model considered, when BS uses the MRC receiver based on the LMMSE, the achievable rate of user $k$ can be approximated simply by

\begin{equation}\label{rgu1}
{R_k} \approx \frac{{T - K}}{T}{\log _2}\left( {1 + {{\widehat  {SINR} }_k}} \right).
\end{equation}
In order to see the individual impacts of the LOS component and the Rayleigh component on ${\widehat  {SINR} }_k$, we divide the signal power and the interfering power into two parts respectively as follows:
\begin{equation}\label{eq:sinry}{\widehat  {SINR} _k} = \frac{{{S_{LOS,k}} + {S_{Ray,k}}}}{{{I_{LOS,k}} + {I_{Ray,k}}}},\end{equation}
\[{S_{LOS,k}} = \frac{{\lambda _{1,k}^2}}{{{{\left( {{\vartheta _k} + 1} \right)}^2}}}\left[ {2\frac{{{\vartheta _k}}}{{{\vartheta _k} + 1}}\left( {N\lambda _{1,k}^{}\sum\limits_{n = 1}^N {\delta _{k,n}^2}  + {\bm{\bar g}}_{1,k}^{\rm{H}}{\bm{U\Delta }}_k^2{{\bm{U}}^{\rm{H}}}{{{\bm{\bar g}}}_{1,k}}} \right) + \vartheta _k^2{N^2}} \right],\]
\[{S_{Ray,k}} = \frac{{\lambda _{1,k}^2}}{{{{\left( {{\vartheta _k} + 1} \right)}^2}}}\left\{ {{{\left( {\frac{{{\lambda _{1,k}}}}{{{\vartheta _k} + 1}}} \right)}^2}\left[ {\sum\limits_{n = 1}^N {\delta _{k,n}^4}  + {{\left( {\sum\limits_{n = 1}^N {\delta _{k,n}^2} } \right)}^2}} \right]} \right\},\]
\[{I_{Ray,k}} =  \frac{{\lambda _{1,k}^2}}{{{{\left( {{\vartheta _k} + 1} \right)}^2}}}\left\{ {\sum\limits_{n = 1}^N {{a_n}\delta _{k,n}^2}  + \sum\limits_{l = 2}^L {\lambda _{l,k}^2\left[ {\sum\limits_{n = 1}^N {\delta _{k,n}^4}  + {{\left( {\sum\limits_{n = 1}^N {\delta _{k,n}^2} } \right)}^2}} \right]}  + \frac{{\sum\limits_{n = 1}^N {\delta _{k,n}^2} }}{{{p_u}}}} \right\},\]
\[\begin{array}{l}
{I_{LOS,k}} = \frac{{{\vartheta _k}}}{{{\vartheta _k} + 1}}{\bm{\bar g}}_{1,k}^{\rm{H}}\left[ {{\bm{UB}}{{\bm{U}}^{\rm{H}}} + \sum\limits_{i \ne k}^K {\left( {\frac{{{\vartheta _i}}}{{{\vartheta _i} + 1}}{{{\bm{\bar g}}}_{1,i}}{\bm{\bar g}}_{1,i}^{\rm{H}}} \right)} } \right]{{{\bm{\bar g}}}_{1,k}}\\
\quad \quad \quad  \quad \quad + \frac{1}{{{{\left( {{\vartheta _k} + 1} \right)}^2}}}\left[ {\lambda _{1,k}^2\sum\limits_{i \ne k}^K {\left( {\frac{{{\vartheta _i}}}{{{\vartheta _i} + 1}}{\bm{\bar g}}_{1,i}^{\rm{H}}{\bm{U\Delta }}_k^2{{\bm{U}}^{\rm{H}}}{{{\bm{\bar g}}}_{1,i}}} \right)}  + \frac{1}{{{p_u}}}{\vartheta _k}\left( {{\vartheta _k} + 1} \right)N{\lambda _{1,k}}} \right].
\end{array}\]
where ${\bm{R}} = {\bm{UD}}{{\bm{U}}^{\rm{H}}}$, with ${\bm{D}} = \text{diag}\left( {{d_1},{d_2}, \cdots ,{d_N}} \right)$. And ${\bm{B}} \buildrel \Delta \over = {\bm{{\rm A}}} + {\bm{\Delta }}_k^2\sum\limits_{l = 2}^L {\lambda _{l,k}^2} $, where ${{\bm{\Delta }}_k} = \text{diag}\left( {{\delta _{k,1}}, \cdots ,{\delta _{k,N}}} \right)$ is a diagonal matrix with element
\begin{equation}\label{eq:delta}
{\delta _{k,n}} = \frac{{{d_n}}}{{\sqrt {\frac{{{\lambda _{1,k}}{d_n}}}{{{\vartheta _k} + 1}} + \sum\limits_{l = 2}^L {{\lambda _{l,k}}{d_n}}  + \frac{1}{{{p_P}}}} }},
\end{equation}
and
\begin{align}
{\bm{A}}&=\text{diag}\left( {{a_1},{a_2}, \cdots ,{a_N}} \right)\nonumber \\
 &= \sum\limits_{i = 1}^K {\left( {\frac{{{\lambda _{1,i}}{\bm{D}}}}{{{\vartheta _i} + 1}}} \right)}  + \sum\limits_{l = 2}^L {\sum\limits_{i = 1}^K {{\lambda _{l,i}}{\bm{D}}} }  - {\left( {\frac{{{\lambda _{1,k}}}}{{{\vartheta _k} + 1}}} \right)^2}{{\bm{\Delta }}_k^2} - \sum\limits_{l = 2}^L {\lambda _{l,k}^2{{\bm{\Delta }}_k^2}}.
 \end{align}
\end{theorem}

\begin{IEEEproof}
See Appendix  $\text{\ref{scell_dl_1}}$.
\end{IEEEproof}

\begin{theorem}\label{thm:lmmsei}
 When the BS antenna number is very large, the achievable rate of user $k$ in the reference cell can be approximated by
\begin{equation}\label{rgu2}
{R_k} \to \frac{{T - K}}{T}{\log _2}\left( {1 + {{\widehat  {SINR} }_k^{\infty}}} \right),
\end{equation}
where
\begin{equation}\label{form2}
{{\widehat  {SINR} }_k^{\infty}}= \frac{{{{\left( {\frac{{{\lambda _{1,k}}}}{{{\vartheta _k} + 1}} + \frac{{{\vartheta _k}}}{{{{\left( {\sum\limits_{n = 1}^N {\delta _{k,n}^2} } \right)} \mathord{\left/
 {\vphantom {{\left( {\sum\limits_{n = 1}^N {\delta _{k,n}^2} } \right)} N}} \right.
 \kern-\nulldelimiterspace} N}}}} \right)}^2}}}{{\sum\limits_{l = 2}^L {\lambda _{l,k}^2} }}.
\end{equation}
\end{theorem}

\begin{IEEEproof}
See Appendix  $\text{\ref{mcell_dl_1}}$.
\end{IEEEproof}

Looking into the proof of Theorem \ref{thm:lmmsei}, we find that as the number of BS antennas is very large, the effects of uncorrelated receiver noise and interferences are eliminated completely, leaving the users using the same pilot as the only interferences. The signal power of the reference user includes LOS part and the Rayleigh part. And when ${\vartheta _k} = 0$,
\[{{\widehat  {SINR} }_k^{\infty}} = \frac{{\lambda _{1,k}^2}}{{\sum\limits_{l = 2}^L {\lambda _{l,k}^2} }},
\]
which coincide with the results in the Rayleigh channel.

We further analyze the impacts of LOS part on ${{\widehat  {SINR} }_k^{\infty}}$. First we make the following definitions,

\[\frac{{{\lambda _{1,k}}}}{{{\vartheta _k} + 1}} + \frac{{{\vartheta _k}}}{{{{\left( {\sum\limits_{n = 1}^N {\frac{{d_n^2}}{{\left( {\frac{{{\lambda _{1,k}}}}{{{\vartheta _k} + 1}} + \sum\limits_{l = 2}^L {{\lambda _{l,k}}} } \right){d_n} + \frac{1}{{{p_P}}}}}} } \right)} \mathord{\left/
 {\vphantom {{\left( {\sum\limits_{n = 1}^N {\frac{{d_i^2}}{{\left( {\frac{{{\lambda _{1,k}}}}{{{\vartheta _k} + 1}} + \sum\limits_{l = 2}^L {{\lambda _{l,k}}} } \right){d_n} + \frac{1}{{{p_P}}}}}} } \right)} N}} \right.
 \kern-\nulldelimiterspace} N}}} \buildrel \Delta \over = x,\]
 \[\frac{{{\vartheta _k}}}{{{{\left( {\sum\limits_{n = 1}^N {\frac{{d_n^2}}{{\left( {\frac{{{\lambda _{1,k}}}}{{{\vartheta _k} + 1}} + \sum\limits_{l = 2}^L {{\lambda _{l,k}}} } \right){d_n} + \frac{1}{{{p_P}}}}}} } \right)} \mathord{\left/
 {\vphantom {{\left( {\sum\limits_{n = 1}^N {\frac{{d_i^2}}{{\left( {\frac{{{\lambda _{1,k}}}}{{{\vartheta _k} + 1}} + \sum\limits_{l = 2}^L {{\lambda _{l,k}}} } \right){d_n} + \frac{1}{{{p_P}}}}}} } \right)} N}} \right.
 \kern-\nulldelimiterspace} N}}} \buildrel \Delta \over = y.\]
After doing some algebraic operation to $y$, we get
\[y{\rm{ = }}\frac{{\rm{1}}}{{\sum\limits_{n = 1}^N {\frac{{{{d_n^2} \mathord{\left/
 {\vphantom {{d_n^2} N}} \right.
 \kern-\nulldelimiterspace} N}}}{{\frac{{{\vartheta _k}{\lambda _{1,k}}{d_n}}}{{{\vartheta _k} + 1}} + {\vartheta _k}\left( {\sum\limits_{l = 2}^L {{\lambda _{l,k}}} {d_n} + \frac{1}{{{p_P}}}} \right)}}} }},\]
which shows that with the increase of ${\vartheta _k}$,  $\frac{{{\vartheta _k}}}{{{\vartheta _k} + 1}}$ increases, so that $y$ increases. And the increase of $y$ consists of the increase of $\frac{{{\vartheta _k}{\lambda _{1,k}}{d_n}}}{{{\vartheta _k} + 1}}$ and ${\vartheta _k}\left( {\sum\limits_{l = 2}^L {{\lambda _{l,k}}} {d_n} + \frac{1}{{{p_P}}}} \right)$. Since
\[\frac{{\rm{1}}}{{\sum\limits_{n = 1}^N {\frac{{{{d_n^2} \mathord{\left/
 {\vphantom {{d_n^2} N}} \right.
 \kern-\nulldelimiterspace} N}}}{{\frac{{{\vartheta _k}{\lambda _{1,k}}{d_n}}}{{{\vartheta _k} + 1}} + {\vartheta _k}\left( {\sum\limits_{l = 2}^L {{\lambda _{l,k}}} {d_n} + \frac{1}{{{p_P}}}} \right)}}} }} \ge \frac{{\rm{1}}}{{\sum\limits_{n = 1}^N {\frac{{{{d_n^2} \mathord{\left/
 {\vphantom {{d_n^2} N}} \right.
 \kern-\nulldelimiterspace} N}}}{{\frac{{{\vartheta _k}{\lambda _{1,k}}{d_n}}}{{{\vartheta _k} + 1}}}}} }}{\rm{ = }}\frac{{{\vartheta _k}{\lambda _{1,k}}}}{{{\vartheta _k} + 1}},\]
 and $\frac{{{\lambda _{1,k}}}}{{{\vartheta _k} + 1}} + \frac{{{\vartheta _k}{\lambda _{1,k}}}}{{{\vartheta _k} + 1}}{\rm{ = }}{\lambda _{1,k}}$ is independent of ${\vartheta _k}$, then we can make the conclusion that $x$ increases with the increase of ${\vartheta _k}$, which means the LOS part raises the user's SINR.

According to \eqref{form2}, the LOS part of the signal power is related to the correlation of the BS antennas, while the Rayleigh part is independent of the correlation. And according to \eqref{eq:inequa1}, we get \[\frac{1}{{\frac{{{\lambda _{1,k}}}}{{{\vartheta _k} + 1}} + \sum\limits_{l = 2}^L {{\lambda _{l,k}}} }} > \frac{{\sum\limits_{n = 1}^N {\delta _{k,n}^2} }}{N} \ge \frac{1}{{\frac{{{\lambda _{1,k}}}}{{{\vartheta _k} + 1}} + \sum\limits_{l = 2}^L {{\lambda _{l,k}}}  + \frac{1}{{{p_P}}}}},\]
so in the case of big $p_P$, the correlation of BS antennas has little influence on the asymptotic rate.
And as ${p_P} \to \infty $, $\delta _{k,n}^2 \to \frac{{{d_n}}}{{\left( {\frac{{{\lambda _{1,k}}}}{{{\vartheta _k} + 1}} + \sum\limits_{l = 2}^L {{\lambda _{l,k}}} } \right)}}$, and $\frac{{\left( {\sum\limits_{n = 1}^N {\delta _{k,n}^2} } \right)}}{N} \to \frac{1}{{\frac{{{\lambda _{1,k}}}}{{{\vartheta _k} + 1}} + \sum\limits_{l = 2}^L {{\lambda _{l,k}}} }}$ which is independent of BS-sided correlation. So we can conclude that with the increase of pilot power, the receive SINR has less and less impact by the BS correlation when $N$ is very large.

When ${\bm{R}} = {\bm{I}}$, after some algebraic manipulations, ${{\widehat  {SINR} }_k^{\infty}}$ can be expressed as:
\begin{equation}{{\widehat  {SINR} }_k^{\infty}} = \frac{{{{\left[ {{\lambda _{1,k}} + {\vartheta _k}\left( {\sum\limits_{l = 2}^L {{\lambda _{l,k}}}  + \frac{1}{{{p_P}}}} \right)} \right]}^2}}}{{\sum\limits_{l = 2}^L {\lambda _{l,k}^2} }},\end{equation}
which coincides with \cite{cmrice}. And on the other hand, the expression confirm our conclusion that the LOS part raises the user's SINR.

\begin{theorem}\label{powersl}
If the transmit power of each user is scaled down as ${p_u} = {E_u}{N^{ - \varepsilon }}$ for a fixed ${E_u}$ and $\varepsilon>0$, when the number of antennas increases, the user's uplink rate of MRC receiver based on pilot assisted LMMSE channel estimate approaches
\begin{equation}\label{rgu3}
{R_k} \to \frac{{T - K}}{T}{\log _2}\left( {1 + {{\widehat  {SINR} }_k^{ps}}} \right),
\end{equation}
where
\begin{equation}\label{form3}
{{\widehat  {SINR} }_k^{ps}} = \frac{{{N^{1 - \varepsilon }}{E_u}{\lambda _{1,k}}{\vartheta _k}}}{{\left( {{\vartheta _k} + 1} \right)}},
\end{equation}
and for ${\vartheta _k} = 0$, \begin{equation}\label{form0}SINR_k^{ps} = \frac{{\lambda _{1,k}^2}}{{\sum\limits_{l = 2}^L {\lambda _{l,k}^2}  + \frac{1}{{KE_u^2{N^{ - 2\varepsilon }}\sum\limits_{n = 1}^N {d_n^2} }}}}.\end{equation}
\end{theorem}

\begin{IEEEproof}
See Appendix  $\text{\ref{z4}}$.
\end{IEEEproof}

We can make several observations from Theorem \ref{powersl}, for Rayleigh fading, the SINR is dependent on the BS-sided correlation when the power scaling is taken (however, the influence is limited). And $\varepsilon$ should be no more than 1/2 to obtain a non-zero constant value with increasing $N$; for Ricean fading, the SINR is independent of the BS-sided correlation when the power scaling is taken. And $\varepsilon$ should be no more than 1 to obtain a non-zero constant value with increasing $N$. Comparing \eqref{form3} and \eqref{form0}, we find that for Ricean fading, the uplink rate only depends on the LOS related power, while the Rayleigh related power and the interference from users in other cells disappear.

\section{Uplink rate analysis using LOS component as channel estimate}

The received signal $\bm{y}$ can be expressed in the following form:
\[{\bm{y}} = \sqrt {{p_u}} \sum\limits_{i = 1}^K {\frac{{\sqrt {{\vartheta _i}} }}{{\sqrt {{\vartheta _i} + 1} }}{{{\bm{\bar g}}}_{1,i}}} {x_{1,i}} + \sqrt {{p_u}} \sum\limits_{i = 1}^K {\frac{{{{{\bm{\mathord{\buildrel{\lower3pt\hbox{$\scriptscriptstyle\frown$}}
\over g} }}}_{1,i}}{x_{1,i}}}}{{\sqrt {{\vartheta _i} + 1} }}}  + \sqrt {{p_u}} \sum\limits_{l = 2}^L {\sum\limits_{i = 1}^K {{{\bm{g}}_{l,i}}} {x_{l,i}}}  + {\bm{w}}.\]

As mentioned above, both the deterministic LOS component and Ricean factor matrix ${\bm{\Omega }}$ are perfectly known. In this part we take the LOS component as channel estimate while taking the scattered part as interference. Let ${\bm{z}} = \sqrt {{p_u}} \sum\limits_{i = 1}^K {\frac{{{{{\bm{\mathord{\buildrel{\lower3pt\hbox{$\scriptscriptstyle\frown$}}
\over g} }}}_{1,i}}{x_{1,i}}}}{{\sqrt {{\vartheta _i} + 1} }}}  + \sqrt {{p_u}} \sum\limits_{l = 2}^L {\sum\limits_{i = 1}^K {{{\bm{g}}_{l,i}}} {x_{l,i}}}  + {\bm{w}}$ identified as the effective noise with covariance
\[{\bm{\Omega }} = E\left( {{\bm{z}}{{\bm{z}}^{\rm{H}}}} \right) = {p_u}\sum\limits_{i = 1}^K {\frac{{{\lambda _{1,i}}}}{{{\vartheta _i} + 1}}}  + {p_u}\sum\limits_{l = 2}^L {\sum\limits_{i = 1}^K {{\lambda _{l,k}}} {\bm{R}}}  + {{\bm{I}}_N}.\]

Based on the assumptions, the MRC filter ${{\bm{\bar c}}_k} = {\bm{\bar g}}_{1,k}^{\rm{H}}$. And after MRC filtering, the signal of the $k$th user is
\[{r_k} = \sqrt {{p_u}} \frac{{\sqrt {{\vartheta _k}} {\bm{\bar g}}_{1,k}^{\rm{H}}{{{\bm{\bar g}}}_{1,k}}{x_{1,k}}}}{{\sqrt {{\vartheta _k} + 1} }} + \frac{{{\bm{\bar g}}_{1,k}^{\rm{H}}{{{\bm{\mathord{\buildrel{\lower3pt\hbox{$\scriptscriptstyle\frown$}}
\over g} }}}_{1,k}}{x_{1,k}}}}{{\sqrt {{\vartheta _k} + 1} }} + \sqrt {{p_u}} \sum\limits_{i \ne k}^K {{\bm{\bar g}}_{1,k}^{\rm{H}}{{\bm{g}}_{1,i}}{x_{1,i}}}  + \sqrt {{p_u}} \sum\limits_{l = 2}^L {\sum\limits_{i = 1}^K {{\bm{\bar g}}_{1,k}^{\rm{H}}{{\bm{g}}_{l,i}}} {x_{l,i}}}  + {\bm{\bar g}}_{1,k}^{\rm{H}}{\bm{w}}.\]

According to the worst case uncorrelated additive noise Theorem in  \cite{BHassibitraining}, the lower bound of the $k$th user's achievable rate is
\[{R_k} = {\log _2}\left( {1 + \frac{{\frac{{{\vartheta _k}}}{{{\vartheta _k} + 1}}{{\left| {{\bm{\bar g}}_{1,k}^{\rm{H}}{{{\bm{\bar g}}}_{1,k}}} \right|}^2}}}{{E\left[ {{\bm{\bar g}}_{1,k}^{\rm{H}}\left( {\frac{{{{{\bm{\mathord{\buildrel{\lower3pt\hbox{$\scriptscriptstyle\frown$}}
\over g} }}}_{1,k}}{\bm{\mathord{\buildrel{\lower3pt\hbox{$\scriptscriptstyle\frown$}}
\over g} }}_{1,k}^{\rm{H}}}}{{{\vartheta _k} + 1}} + \sum\limits_{i \ne k}^K {{{\bm{g}}_{1,i}}{\bm{g}}_{1,i}^{\rm{H}}}  + \sum\limits_{l = 2}^L {\sum\limits_{i = 1}^K {{{\bm{g}}_{l,i}}} {\bm{g}}_{l,i}^{\rm{H}}}  + \frac{{{{\bm{I}}_N}}}{{{p_u}}}} \right){{{\bm{\bar g}}}_{1,k}}} \right]}}} \right).\]

\begin{theorem}\label{thm:lose}
For system model considered, when BS uses the MRC receiver based on the LOS component, the lower bound of the achievable rate of user $k$ is
\begin{equation}\label{eq:losifc}{R_k} = {\log _2}\left( {1 + {\overline {SINR} _k}} \right),\end{equation}
where
\begin{equation}\label{eq:losif}{\overline {SINR} _k} = \frac{{\frac{{{\vartheta _k}}}{{{\vartheta _k} + 1}}{{\left( {{\lambda _{1,k}}N} \right)}^2}}}{{ {\lambda _{1,k}}\sum\limits_{i \ne k}^K {\frac{{{\vartheta _i}{\lambda _{1,i}}}}{{{\vartheta _i} + 1}}{{\left| {{\rho _{k,i}}} \right|}^2}}  + \sum\limits_{i =1}^K {\frac{{{\lambda _{1,i}}{\bm{\bar g}}_{1,k}^{\rm{H}}{\bm{R}}{{{\bm{\bar g}}}_{1,k}}}}{{{\vartheta _i} + 1}}}  + \sum\limits_{l = 2}^L {\sum\limits_{i = 1}^K {{\lambda _{l,i}}{\bm{\bar g}}_{1,k}^{\rm{H}}{\bm{R}}{{{\bm{\bar g}}}_{1,k}}} }  + \frac{{{\lambda _{1,k}}N}}{{{p_u}}}}}.\end{equation}
\end{theorem}

\begin{IEEEproof}
Define \begin{equation}\label{eq:los}{\overline {SINR} _k} \buildrel \Delta \over = \frac{{\frac{{{\vartheta _k}}}{{{\vartheta _k} + 1}}{{\left| {{\bm{\bar g}}_{1,k}^{\rm{H}}{{{\bm{\bar g}}}_{1,k}}} \right|}^2}}}{{E\left[ {{\bm{\bar g}}_{1,k}^{\rm{H}}\left( {\frac{{{{{\bm{\mathord{\buildrel{\lower3pt\hbox{$\scriptscriptstyle\frown$}}
\over g} }}}_{1,k}}{\bm{\mathord{\buildrel{\lower3pt\hbox{$\scriptscriptstyle\frown$}}
\over g} }}_{1,k}^{\rm{H}}}}{{{\vartheta _k} + 1}} + \sum\limits_{i \ne k}^K {{{\bm{g}}_{1,i}}{\bm{g}}_{1,i}^{\rm{H}}}  + \sum\limits_{l = 2}^L {\sum\limits_{i = 1}^K {{{\bm{g}}_{l,i}}} {\bm{g}}_{l,i}^{\rm{H}}}  + \frac{{{{\bm{I}}_N}}}{{{p_u}}}} \right){{{\bm{\bar g}}}_{1,k}}} \right]}},\end{equation}
and each element of the numerator and denominator can be simplified as follows:
\[{\left| {{\bm{\bar g}}_{1,k}^{\rm{H}}{{{\bm{\bar g}}}_{1,k}}} \right|^2} = {\left( {{\lambda _{1,k}}N} \right)^2},\]
\[E\left[ {{\bm{\bar g}}_{1,k}^{\rm{H}}\left( {\frac{{{{\bm{I}}_N}}}{{{p_u}}}} \right){{{\bm{\bar g}}}_{1,k}}} \right] = \frac{{{\lambda _{1,k}}N}}{{{p_u}}},\]
\[E\left[ {{\bm{\bar g}}_{1,k}^{\rm{H}}\frac{{{{{\bm{\mathord{\buildrel{\lower3pt\hbox{$\scriptscriptstyle\frown$}}
\over g} }}}_{1,k}}{\bm{\mathord{\buildrel{\lower3pt\hbox{$\scriptscriptstyle\frown$}}
\over g} }}_{1,k}^{\rm{H}}}}{{{\vartheta _k} + 1}}{{{\bm{\bar g}}}_{1,k}}} \right] = \frac{{\lambda _{1,k}^{}{\bm{\bar g}}_{1,k}^{\rm{H}}{\bm{R}}{{{\bm{\bar g}}}_{1,k}}}}{{{\vartheta _k} + 1}},\]
\[E\left[ {{\bm{\bar g}}_{1,k}^{\rm{H}}\left( {\sum\limits_{l = 2}^L {\sum\limits_{i = 1}^K {{{\bm{g}}_{l,i}}} {\bm{g}}_{l,i}^{\rm{H}}} } \right){{{\bm{\bar g}}}_{1,k}}} \right] = \sum\limits_{l = 2}^L {\sum\limits_{i = 1}^K {{\lambda _{l,i}}{\bm{\bar g}}_{1,k}^{\rm{H}}{\bm{R}}{{{\bm{\bar g}}}_{1,k}}} } \]
\[E\left[ {{\bm{\bar g}}_{1,k}^{\rm{H}}\left( {\sum\limits_{i \ne k}^K {{{\bm{g}}_{1,i}}{\bm{g}}_{1,i}^{\rm{H}}} } \right){{{\bm{\bar g}}}_{1,k}}} \right] = {\lambda _{1,k}}\sum\limits_{i \ne k}^K {\frac{{{\vartheta _i}{\lambda _{1,i}}}}{{{\vartheta _i} + 1}}{{\left| {{\rho _{k,i}}} \right|}^2}}  + \sum\limits_{i \ne k}^K {\frac{{{\lambda _{1,i}}{\bm{\bar g}}_{1,k}^{\rm{H}}{\bm{R}}{{{\bm{\bar g}}}_{1,k}}}}{{{\vartheta _i} + 1}}}, \]
then substituting all the expressions into \eqref{eq:los}, Theorem \ref{thm:lose} is proved.
\end {IEEEproof}

Similarly, we give the expression of $SINR_k^{LOS}$ when the number of BS antennas $N$ is very large.

\begin{theorem}\label{thm:losin}
Using LOS component as channel estimate, as the number of BS antennas $N$ is very large, the achievable rate of user $k$ can be approximated as follow
\begin{equation}\label{eq:rlos}
{R_k} \to {\log _2}\left( {1 + {\overline {SINR} _k^{\infty}}} \right),
\end{equation}
where
\begin{equation}\label{eq:losxab}{\overline {SINR} _k^{\infty}} = \frac{{\frac{{{\vartheta _k}}}{{{\vartheta _k} + 1}}\left( {{\lambda _{1,k}}N} \right)}}{{\sum\limits_{i = 1}^K {\frac{{{\lambda _{1,i}}{\bm{\bar g}}_{1,k}^{\rm{H}}{\bm{R}}{{{\bm{\bar g}}}_{1,k}}}}{{N{\lambda _{1,k}}\left( {{\vartheta _i} + 1} \right)}}}  + \sum\limits_{l = 2}^L {\sum\limits_{i = 1}^K {\frac{{{\lambda _{l,i}}{\bm{\bar g}}_{1,k}^{\rm{H}}{\bm{R}}{{{\bm{\bar g}}}_{1,k}}}}{{N{\lambda _{1,k}}}}} }  + \frac{1}{{{p_u}}}}}.\end{equation}
\end{theorem}

\begin{IEEEproof}
Since ${d _{\min }}N{\lambda _{1,k}} \le {\bm{\bar g}}_{1,k}^{\rm{H}}{\bm{R}}{{\bm{\bar g}}_{1,k}} \le {d _{\max }}N{\lambda _{1,k}}$, we define ${\bm{\bar g}}_{1,k}^{\rm{H}}{\bm{R}}{{\bm{\bar g}}_{1,k}}=\alpha {\lambda _{l,k}}N$, $\alpha >0$ and limited, which depends the BS-sided correlation. So ${\frac{{{\bm{\bar g}}_{1,k}^{\rm{H}}{\bm{R}}{{{\bm{\bar g}}}_{1,k}}}}{{N}}} $ is a constant depending on the BS-sided correlation. Besides, ${\rho _{k,i}} = \frac{{1 - {e^{jN{\varphi _{ki}}}}}}{{1 - {e^{j{\varphi _{ki}}}}}}$, ${\varphi _{ki}} = \frac{{2\pi d}}{\lambda }\left( {\sin {\theta _k} - \sin {\theta _i}} \right)$
and ${\left| {1 - {e^{jN{\varphi _{ki}}}}} \right|}$ are limited. So when $N\to\infty$, $\frac{{{{\left| {{\rho _{k,i}}} \right|}^2}}}{N} \to 0$. Finally, Theorem \ref{thm:losin} is proved.
\end {IEEEproof}

Theorem \ref{thm:losin} shows that the power related to LOS component increases linearly with the increase of the number of BS antennas, while the power related to the Rayleigh component of reference cell and the interfering cell has nothing to do with the number of BS antennas. Based on these, we can easily speculate that as the number of BS antennas grows bigger and bigger, the user's rate using LOS component will gradually exceed the rate using LMMSE estimate. What's more, according to \eqref{eq:losxab} the rate increases with the increase of $\vartheta$.

For $\bm{R}=\bm{I}$, \eqref{eq:losxab} becomes
\begin{equation}\label{eq:losxabiid}{\overline {SINR} _k^{\infty}}= \frac{{\frac{{{\vartheta _k}}}{{{\vartheta _k} + 1}}\left( {{\lambda _{1,k}}N} \right)}}{{\sum\limits_{i = 1}^K {\frac{{{\lambda _{1,i}}}}{{\left( {{\vartheta _i} + 1} \right)}}}  + \sum\limits_{l = 2}^L {\sum\limits_{i = 1}^K {{\lambda _{l,i}}} }  + \frac{1}{{{p_u}}}}}.\end{equation}
\eqref{eq:losxabiid} shows clearly that with the increase of the number of BS antennas, the user's SINR of the reference cell increases linearly with the number of BS antennas when taking the LOS component as channel estimate.

\begin{theorem}\label{losps}
If the transmit power of each user is scaled down as ${p_u} = {E_u}{N^{ - \varepsilon }}$ for a fixed ${E_u}$ and $\varepsilon>0$, when the number of antennas increases, user's uplink rate of MRC receiver based on the LOS component as channel estimate approaches
\begin{equation}\label{eq:rlosp}
{R_k} \to {\log _2}\left( {1 + {\overline {SINR} _k^{ps}}} \right),
\end{equation}
where
\begin{equation}\label{eq:rlosps}\overline {SINR} _k^{ps} = \frac{{{\vartheta _k}{\lambda _{1,k}}{E_u}}}{{{\vartheta _k} + 1}}{N^{1 - \varepsilon }}.\end{equation}
\end{theorem}

\begin{IEEEproof}
Subsituting ${p_u} = {E_u}{N^{ - \varepsilon }}$ into \eqref{eq:losif},
\[SINR_k^{LOS} = \frac{{\frac{{{\vartheta _k}}}{{{\vartheta _k} + 1}}{{\left( {{\lambda _{1,k}}N} \right)}^2}}}{{\frac{{\lambda _{1,k}^{}{\bm{\bar g}}_{1,k}^{\rm{H}}{\bm{R}}{{{\bm{\bar g}}}_{1,k}}}}{{{\vartheta _k} + 1}} + {\lambda _{1,k}}\sum\limits_{i \ne k}^K {\frac{{{\vartheta _i}{\lambda _{1,i}}}}{{{\vartheta _i} + 1}}{{\left| {{\rho _{k,i}}} \right|}^2}}  + \sum\limits_{i \ne k}^K {\frac{{{\lambda _{1,i}}{\bm{\bar g}}_{1,k}^{\rm{H}}{\bm{R}}{{{\bm{\bar g}}}_{1,k}}}}{{{\vartheta _i} + 1}}}  + \sum\limits_{l = 2}^L {\sum\limits_{i = 1}^K {{\lambda _{l,i}}{\bm{\bar g}}_{1,k}^{\rm{H}}{\bm{R}}{{{\bm{\bar g}}}_{1,k}}} }  + \frac{{{\lambda _{1,k}}{N^{1 + \varepsilon }}}}{{{E_u}}}}}\]
similar to the proof of Theorem \ref{thm:losin}, when $N \to \infty $, \[SINR_k^{LOS} \to \frac{{{\vartheta _k}{\lambda _{1,k}}{E_u}}}{{{\vartheta _k} + 1}}{N^{1 - \varepsilon }} \buildrel \Delta \over =\overline {SINR} _k^{ps}. \]
Thus Theorem \ref{losps} is proved.
\end{IEEEproof}

We find that \eqref{eq:rlosps} is the same as \eqref{form3}. So we conclude that for Ricean channel, when power scaling is taken, pilot contamination will gradually disappear using the pilot assisted LMMSE channel estimation; and the user's SINR will approach to the one using LOS component as channel estimation. So for massive MIMO it has little meaning for pilot-assisted LMMSE estimation.

\section{Numerical Results}
In this section, we validate the analyses presented above through a set of Monte-Carlo simulations. Same as \cite{Wdm ICC}, a 7-cell hexagonal system layout is adopted. The inner cell radius is normalized to one, the distance between two adjacent cells is normalized to 2, and we assume a distance-based path loss model with path loss exponent  $\alpha = 3.7$. To allow for reproducibility of our results, we distribute $K = 10$ users uniformly on a circle of radius 2/3 around each BS according to the random position distribution model in \cite{WdmICC}. The entries of the BS-sided correlation matrix  are modeled via the common exponential correlation model ${\left[ {{{\bm{R}}_{l,i}}} \right]_{m,n}} = {\kappa ^{\left| {m - n} \right|}}$  with  $\kappa $ being the correlation coefficient. Assuming all of users in the reference cell have the identical Ricean factor, and the ratio of the antenna spacing to wavelength is set to 0.5, and unless otherwise stated, the arrival angles are uniformly distributed in the interval $\left. {\left[ { - \frac{\pi }{2},\frac{\pi }{2}} \right.} \right)$ which means ${\theta _k}{\rm{ = }}\frac{{\pi \left( {k - 1} \right)}}{K} - \frac{\pi }{2},k = 1, \cdots ,K$. And the coherence time of the channel is chosen as $T=196$ according to LTE standard.


\begin{figure}
\centering
\includegraphics[scale=.5]{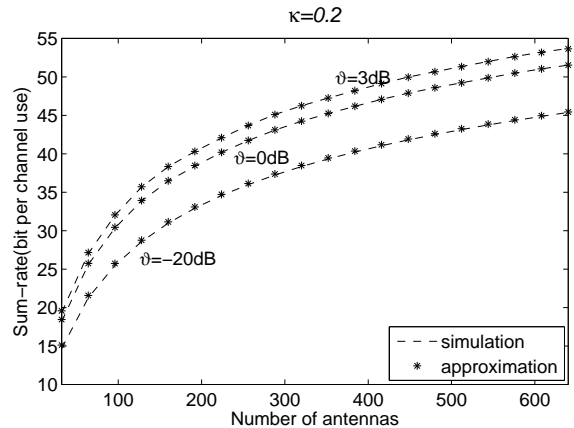}
\caption{Uplink sum-rate as a function of the number of BS antennas $N$ for various Ricean factors
while the pilot assisted LMMSE estimation is used with $\kappa =0.2$.}
\label{Fig5.2}
\end{figure}

\begin{figure}
\centering
\includegraphics[scale=.5]{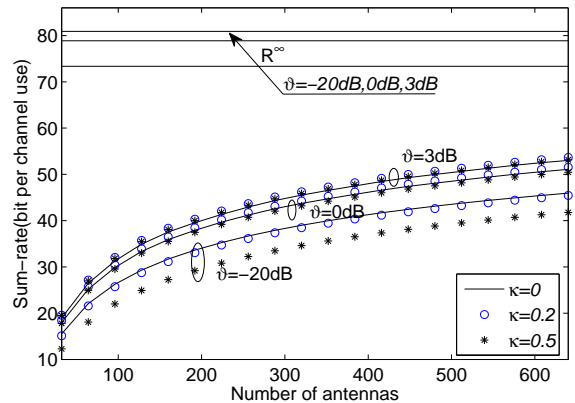}
\caption{Uplink sum-rate as a function of the number of BS antennas $N$ for various Ricean factors and various correlation coefficients while  the pilot assisted LMMSE estimation is used.}
\label{Fig5.3}
\end{figure}


\begin{figure}
\centering
\includegraphics[scale=.5]{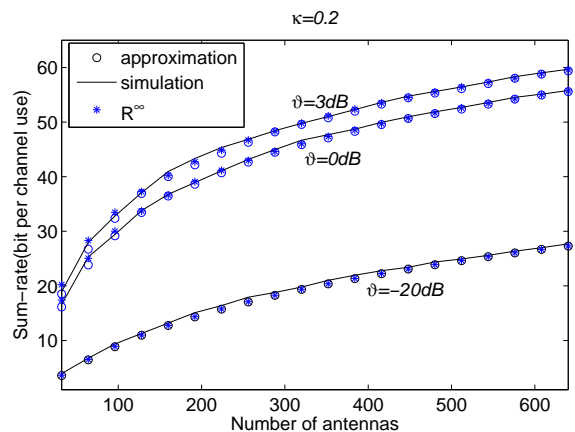}
\caption{Uplink sum-rate as a function of the number of BS antennas $N$ for various Ricean factors while the LOS component is used as channel  estimation with $\kappa =0.2$.}
\label{Fig5.8}
\end{figure}

Here, we show the comparison between the two methods of channel estimation. First of all, we assess the validity of the proposed approximate formulas. Assuming the data power $p_u=10dB$, and the pilot power $p_P=10p_u$, Fig.\ref{Fig5.2} plots the achievable sum-rate as a function of the number of BS antennas $N$ for various Ricean factors while the pilot assisted LMMSE estimation is used with $\kappa=0.2$. Obviously, the approximate expression is quite tight, especially at large $N$. Therefore, in the following, we will use the approximate expression to replace the exact one for the performance analysis. As expected, Fig.\ref{Fig5.2} shows the achievable rate increases with the increase of Ricean factor ${\vartheta _k}$, which is conjectured that the uplink rate with Ricean fading is higher compared to the case of Rayleigh fading. For further analysis of the impact of the BS correlation, Fig.\ref{Fig5.3} plots  the achievable sum-rate as a function of the number of BS antennas $N$ for various Ricean factors and correlation coefficients. $R^\infty$  is the asymptotic sum-rate as $N$ is very large. When $N$ is finite, it can be clearly seen from  Fig.\ref{Fig5.3} that for small ${\vartheta _k}$ (that is near Rayleigh fading) the correlation decreases the uplink rate. However, as ${\vartheta _k}$ increases, the effect of correlation becomes varied--it seems that the correlation makes a little increase. As expected, when $N$ is very large, the infinite rates for different correlation are too identical to be distinguished in the plot, which means that the impact of the BS spacial correlation on the rate can be negligible.

\begin{figure}
\begin{minipage}{0.5\textwidth}
\centerline{\includegraphics[width=\textwidth]{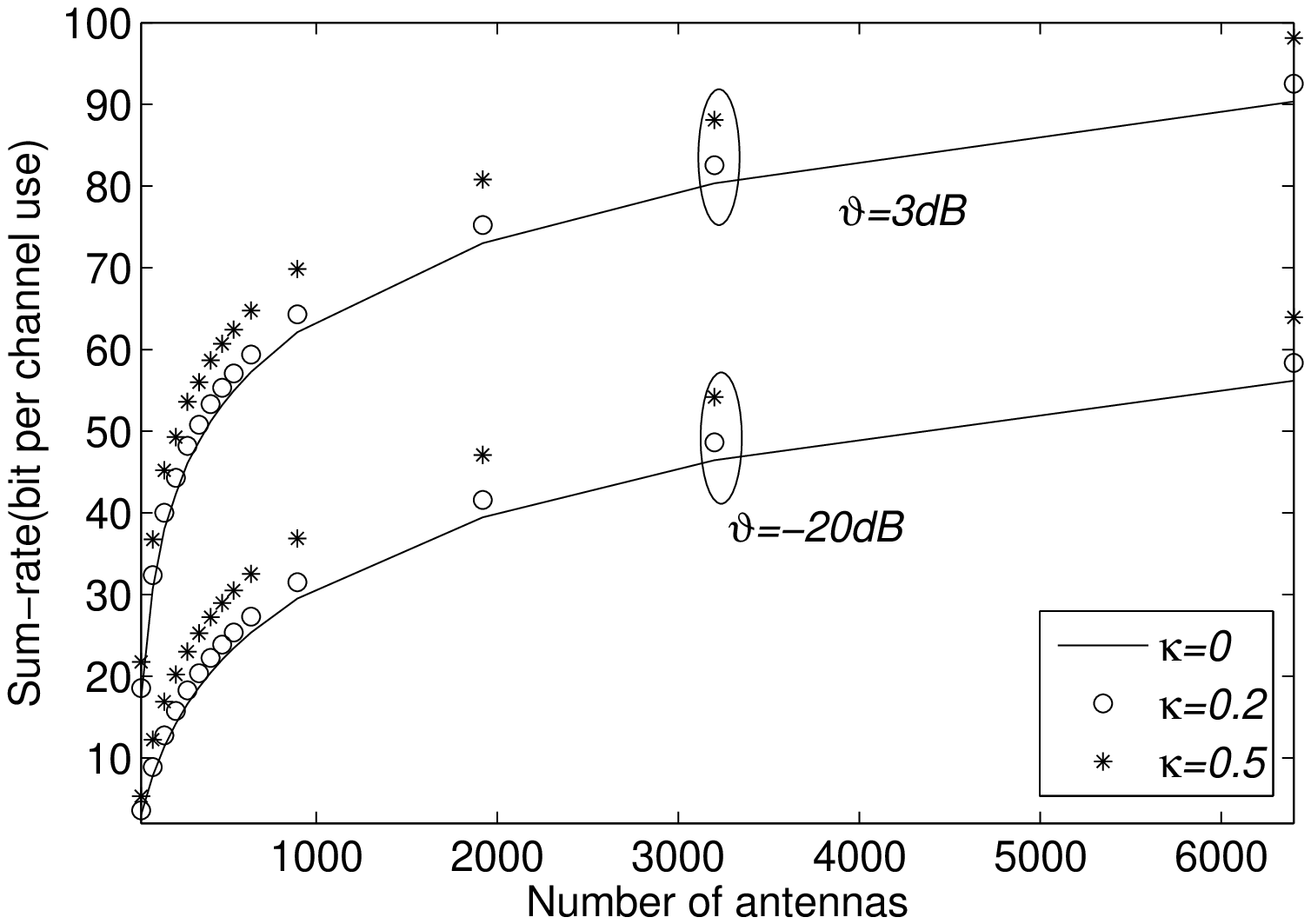}}
\centerline{(a)${\theta _k}{\rm{ = }}\frac{{\pi \left( {k - 1} \right)}}{K} - \frac{\pi }{2},k = 1, \cdots ,K$}
\end{minipage}
\begin{minipage}{0.5\textwidth}
\centerline{\includegraphics[width=\textwidth]{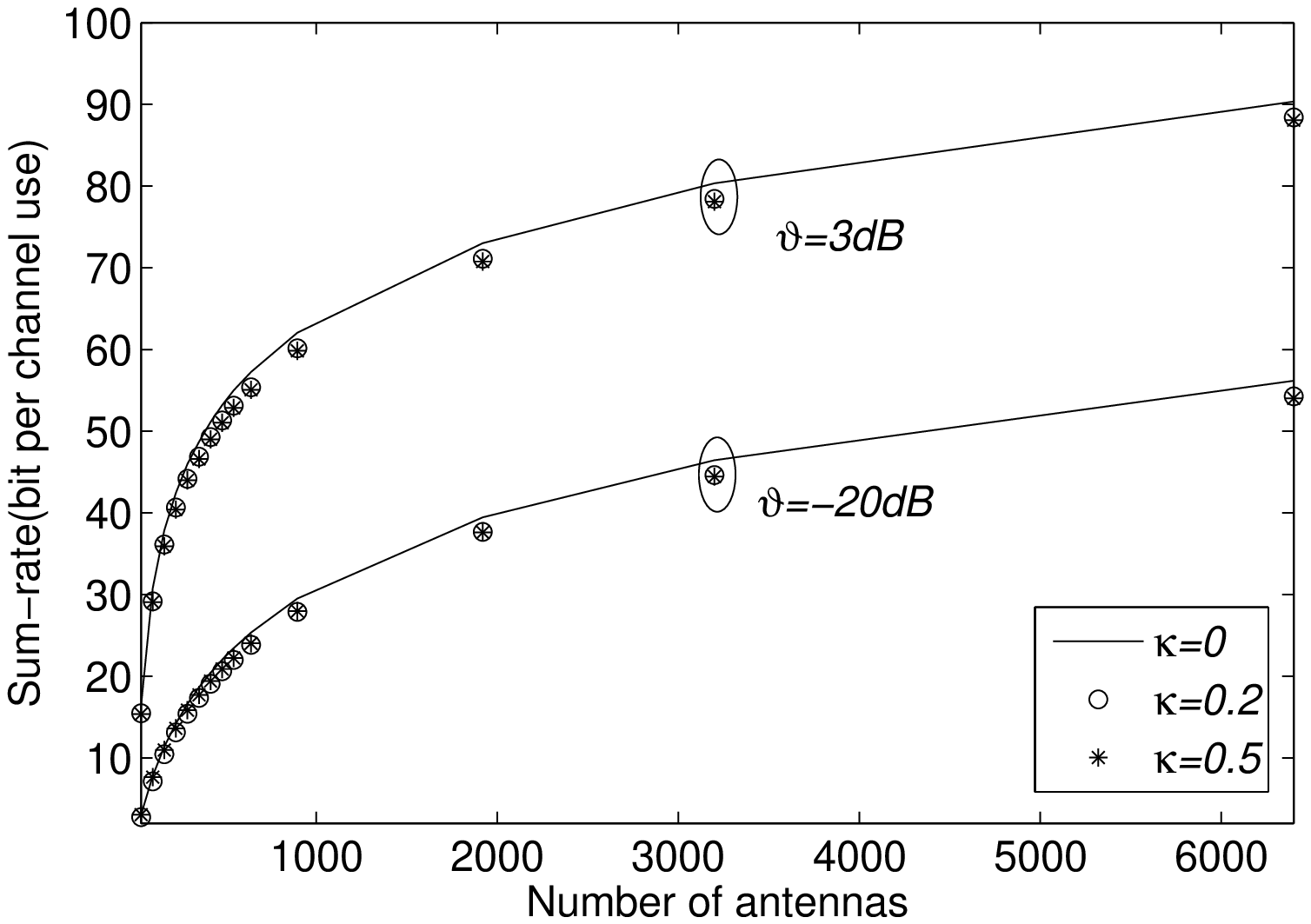}}
\centerline {(b) ${\theta _k}{\rm{ = }}\frac{{{\rm{2k - 1}}}}{{2K}} - \frac{\pi }{4},k = 1, \cdots ,K$}
\end{minipage}
\caption{The impacts of user's position distribution on the uplink sum-rate while the LOS component is used as channel  estimation.}
\label{Fig5.9}
\end{figure}

\begin{figure}
\centering
\includegraphics[scale=.5]{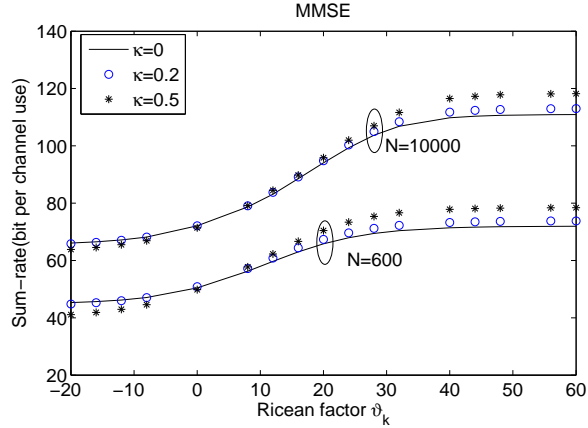}
\caption{The sum-rate against the Ricean factor ${\vartheta _k}$ for different correlation coefficients and numbers of antennas while the pilot assisted LMMSE estimation is used.}
\label{Fig51}
\end{figure}

\begin{figure}
\centering
\includegraphics[scale=.5]{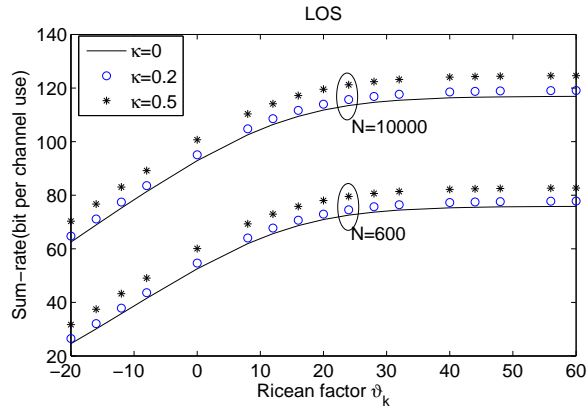}
\caption{The sum-rate against the Ricean factor ${\vartheta _k}$ for different correlation coefficients and numbers of antennas while the LOS component is used as channel estimation.}
\label{Fig52}
\end{figure}

Fig.\ref{Fig5.8} plots the achievable sum-rate as a function of the number of BS antennas $N$ for various Ricean factors while the LOS component is used as channel  estimation with $\kappa=0.2$. Similarly, the approximate expression is quite tight, especially at large $N$. Therefore, in the following, we will use the approximate expression to replace the exact one for the performance analysis. As expected, Fig.\ref{Fig5.8} shows the achievable rate increases with the increase of Ricean factor ${\vartheta _k}$. What's more, we find that \eqref{eq:losifc} approaches the asymptotic  rate \eqref{eq:rlos} when the number of the antennas is not very large. In order to provide an assessment of the influence of spacial correlation, Fig.\ref{Fig5.9}(a) plots the achievable sum-rate as a function of the number of BS antennas $N$ for various Ricean factors and correlation coefficients. It can be clearly seen from Fig.\ref{Fig5.9}(a) that the correlation, on the contrary increases the uplink sum-rate and the increased phenomenon doesn't eliminate with the increase of the number of antennas. With further analysis, we find that the phenomenon is not inevitable, which depends on the position distribution of the users.  For example, if we choose ${\theta _k}{\rm{ = }}\frac{{{\rm{2k - 1}}}}{{2K}} - \frac{\pi }{4},k = 1, \cdots ,K$ , the correlation decreases the uplink sum-rate as shown by Fig.\ref{Fig5.9}(b).

\begin{figure}
\centering
\includegraphics[scale=.5]{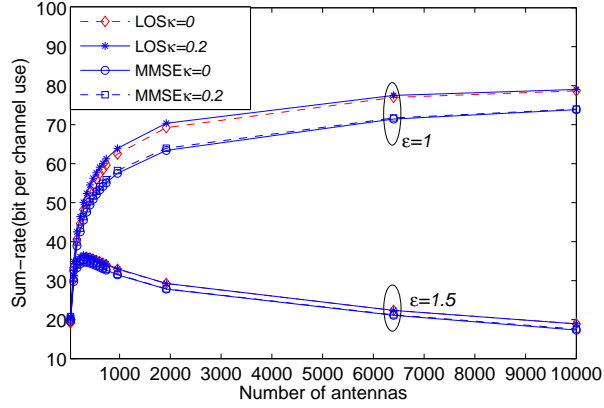}
\caption{Comparison of uplink sum-rate as a function of the number of BS antennas $N$ between pilot assisted LMMSE estimation and LOS component channel as channel estimation while power scaling is taken.}
\label{Fig5.5}
\end{figure}

\begin{figure}
\centering
\includegraphics[scale=.5]{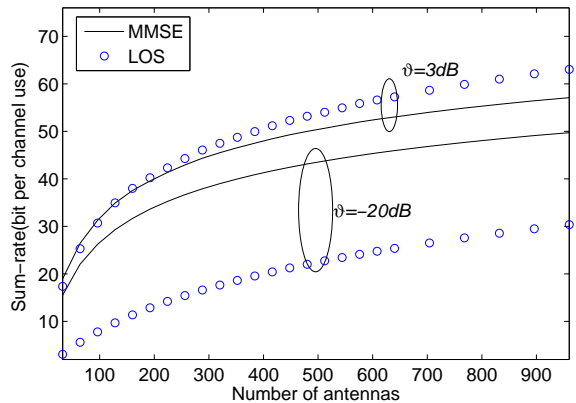}
\caption{Comparison of uplink sum-rate as a function of the number of BS antennas $N$ between pilot assisted LMMSE estimation and LOS component channel as channel estimation.}
\label{Fig5.11}
\end{figure}

For further analysis of the impact of spacial correlation and Ricean factor on the sum-rate, Fig.\ref{Fig51} and Fig.\ref{Fig52} depict the sum-rate against the Ricean factor ${\vartheta _k}$ for different correlation coefficients and numbers of antennas. Analyzing the two figures comprehensively, we can make the following conclusion: for Ricean channel, the spacial correlation can not only decrease but also increase the sum-rate which depends on  distribution of the users' position; For Rayleigh fading channel, the spacing correlation can only decrease the sum-rate. If the specific position distribution causes the LOS component related sum-rate increase, then the impact of the LOS component related sum-rate grows with the increase of Ricean factor, and finally the total sum-rate with pilot assisted LMMSE estimation increases  as shown by Fig.\ref{Fig51}.

Fig.\ref{Fig5.5} validate our power-scaling law in Theorems 3 and 6 with $\vartheta _k=6dB$, $Eu=20dB$. As expected, when the scaling factor $\varepsilon=1 $, sum-rate against the number of antennas shows a trend from rise to be stable. On the other hand, when $\varepsilon=1.5 $, sum-rate against the number of antennas shows a trend from rise to decline, which means the user' power has been reduced too much. And when the transmit power is scaled, the sum-rates of the two methods of channel estimation become gradually identical as $N$ becomes larger (in Fig.\ref{Fig5.5}, $N=1200$). The difference between the two methods is caused by pilot overhead, that is $\frac{{T - K}}{T}$. Fig.\ref{Fig5.5} also shows that the spacial correlation has little influence on the power-scaled sum-rate.

Fig.\ref{Fig5.11} compares the sum-rate against the number of antennas for the two methods of channel estimation. Fig.\ref{Fig5.11} clearly shows that with the increase of $N$, the gap between the two methods gradually gets smaller, and after some specific value (we just call it the turning point) the sum-rate when LOS is used as channel estimation becomes greater than the one of the pilot assisted LMMSE channel estimation. What's more, the bigger ${\vartheta _k}$ is, the smaller the value of the turning point. And in Fig.\ref{Fig5.11}, the turning point is $N=120$ for ${\vartheta _k=3dB}$, $N=300$ for ${\vartheta _k=0dB}$.

\section{Conclusion}

  We studied the performances of spectral efficiency over correlated Ricean fading channel. We deduced the respective analytical expressions for two methods. The first method is MRC based on pilot-assisted LMMSE estimation, the other one is MRC based on LOS part. The impact of Ricean fading has been investigated extensively in the paper. The following conclusions are drawn:

(1) When the BS antenna number is very large, due to the existence of pilot contamination, the asymptotic uplink data rate of pilot-assisted LMMSE estimation method approaches to a finite value which increases with the increase of Ricean factor. However, the asymptotic uplink data rate of LOS method goes linearly with the number of BS antennas. So the uplink achievable rate of LOS method will exceed the one of pilot-assisted LMMSE estimation method with the increase of antenna number.

(2) The expression of the achievable rate of LOS method also showed the correlation between the BS antennas may not only decrease the rate, but also increase the rate, which depends on the locations of the users. So if the locations of the users make antenna correlation increase the rate, with the increase of Ricean factor the rate of pilot-assisted LMMSE estimation method will become larger due to antenna correlation since the effect of LOS part becomes more and more strong.

(3) When the power is scaled the asymptotic expressions of the two methods are the same and both independent of the antenna correlation.

(4) In a word, when the power of LOS part is comparable to the power of fading part for Ricean channel, it has little meaning for pilot-assisted LMMSE estimation for massive MIMO.

\appendices
\section{USEFUL LEMMAS}\label{part:5.1}
\begin{lemma}\label{l1}
If ${X_i},i = 1, \cdots {t_1}$ and ${Y_i},i = 1, \cdots {t_2}$ are both nonnegative random variables, and $X = \sum\limits_{i = 1}^{{t_1}} {{X_i}} $, $Y = \sum\limits_{i = 1}^{{t_2}} {{Y_i}} $, then we get the approximation as follows \cite{cmrice}£¬
\[E\left[ {{{\log }_2}\left( {1 + \frac{X}{Y}} \right)} \right] \approx {\log _2}\left( {1 + \frac{{E\left[ X \right]}}{{E\left[ Y \right]}}} \right)\]
\end{lemma}

\begin{lemma}\label{l2}
If ${Z_1}, \cdots {Z_k}$ are independent, standard random variables, then the sum of their squares, $Q = \sum\limits_{i = 1}^K {Z_i^2} $ is distributed according to the chi-squared distribution with $k$ degrees of freedom, denoted as $Q \sim \chi _k^2$. According to the relationship between the chi-squared distribution and the gamma distribution, we get $Q \sim Gamma\left( {{k \mathord{\left/ {\vphantom {k {2,2}}} \right. \kern-\nulldelimiterspace} {2,2}}} \right)$.
\end{lemma}

\begin{lemma}\label{l3}
If $X$ is the standard circular symmetric complex Gaussian random variable, then ${\left| X \right|^2} \sim Gamma\left( {1,1} \right) $.
\end{lemma}

\begin{lemma}\label{l4}
If $X \sim Gamma\left( {k,\theta } \right)$, then for any $c > 0$, $cX \sim Gamma\left( {k,c\theta } \right)$.
\end{lemma}

\begin{lemma}\label{l5}
If ${X_i} \sim Gamma\left( {{k_i},\theta } \right)$ for $i = 1,2, \cdots ,N$, then $\sum\limits_{i = 1}^N {{X_i}}  \sim Gamma\left( {\sum\limits_{i = 1}^N {{k_i}} ,\theta } \right) $.
\end{lemma}

\begin{lemma}\label{l6}
If ${X_1}, \cdots {X_N}$ are independently distributed, and ${X_i} \sim Gamma\left( {{k_i},{\theta _i}} \right)$, then the first and second moments and variance of $\sum\limits_{i = 1}^N {{X_i}} $ respectively are
\[E\left[ {\sum\limits_{i = 1}^N {{X_i}} } \right] = \sum\limits_{i = 1}^N {{k_i}{\theta _i}} \]
\[E\left[ {{{\left( {\sum\limits_{i = 1}^N {{X_i}} } \right)}^2}} \right] = \sum\limits_{i = 1}^N {{k_i}\theta _i^2}  + {\left( {\sum\limits_{i = 1}^N {{k_i}{\theta _i}} } \right)^2}\]
\[{\mathop{\rm var}} \left[ {\sum\limits_{i = 1}^N {{X_i}} } \right] = \sum\limits_{i = 1}^N {{k_i}\theta _i^2} \]
\end{lemma}

\section{PROOF}
\subsection{Proof of Theorem \ref{thm:lmmse}}\label{scell_dl_1}
\begin{IEEEproof}
We start by decomposing the correlation matrix as
${\bm{R}} = {\bm{UD}}{{\bm{U}}^{\rm{H}}}$, where $\bm{U}$ is a unitary matrix and $\bm{D}$ is a diagonal matrix whose elements on the principal diagonal are the eigenvalues of $\bm{R}$, denoted by $d_1, \cdots , d_N$, then \[\begin{array}{l}
{{\bm{Q}}_k} = {\left( {\frac{{{\lambda _{1,k}}{\bm{UD}}{{\bm{U}}^{\rm{H}}}}}{{{\vartheta _k} + 1}} + \sum\limits_{l = 2}^L {{\lambda _{l,k}}{\bm{UD}}{{\bm{U}}^{\rm{H}}}}  + \frac{{{{\bm{I}}_N}}}{{{p_P}}}} \right)^{ - 1}}\\
\quad \;\; \buildrel \Delta \over = {\bm{UD}}_k^Q{{\bm{U}}^{\rm{H}}}
\end{array}\]
similarly,
\[{\bm{RQ}}_k^{\frac{1}{2}} \buildrel \Delta \over = {\bm{U}}{{\bm{\Delta }}_k}{{\bm{U}}^{\rm{H}}}\]
where ${{\bm{\Delta }}_k} = \text{diag}\left( {{\delta _{k,1}}, \cdots ,{\delta _{k,N}}} \right)$ is a diagonal matrix with $\delta _{k,n}$ expressed as
\[{\delta _{k,n}} = \frac{{{d_n}}}{{\sqrt {\frac{{{\lambda _{1,k}}{d_n}}}{{{\vartheta _k} + 1}} + \sum\limits_{l = 2}^L {{\lambda _{l,k}}{d_n}}  + \frac{1}{{{p_P}}}} }}.\]

1. The power of the desired signal
\[\begin{array}{l}
E\left[ {{{\left| {{\bm{\hat g}}_{1,k}^{\rm{H}}{{{\bm{\hat g}}}_{1,k}}} \right|}^2}} \right]\\
 = E\left[ {{{\left| {{{\left( {\frac{{{\lambda _{1,k}}{\bm{R}}}}{{{\vartheta _k} + 1}}{{\bm{Q}}_k}^{\frac{1}{2}}{{{\bm{\hat h}}}_k} + \frac{{\sqrt {{\vartheta _k}} }}{{\sqrt {{\vartheta _k} + 1} }}{{{\bm{\bar g}}}_{1,k}}} \right)}^{\rm{H}}}\left( {\frac{{{\lambda _{1,k}}{\bm{R}}}}{{{\vartheta _k} + 1}}{{\bm{Q}}_k}^{\frac{1}{2}}{{{\bm{\hat h}}}_k} + \frac{{\sqrt {{\vartheta _k}} }}{{\sqrt {{\vartheta _k} + 1} }}{{{\bm{\bar g}}}_{1,k}}} \right)} \right|}^2}} \right]\\
 = E\left[ {{{\left| {a{\bm{\hat h}}_k^{\rm{H}}{{\bm{Q}}_k}^{\frac{1}{2}}{{\bm{R}}^2}{{\bm{Q}}_k}^{\frac{1}{2}}{{{\bm{\hat h}}}_k} + b{\bm{\bar g}}_{1,k}^{\rm{H}}{{{\bm{\bar g}}}_{1,k}} + c{\bm{\hat h}}_k^{\rm{H}}{{\bm{Q}}_k}^{\frac{1}{2}}{\bm{R}}{{{\bm{\bar g}}}_{1,k}} + c{\bm{\bar g}}_{1,k}^{\rm{H}}{\bm{R}}{{\bm{Q}}_k}^{\frac{1}{2}}{{{\bm{\hat h}}}_k}} \right|}^2}} \right]\\
 = E\left[ {{a^2}{{\left( {{\bm{\hat h}}_k^{\rm{H}}{{\bm{Q}}_k}^{\frac{1}{2}}{{\bm{R}}^2}{{\bm{Q}}_k}^{\frac{1}{2}}{{{\bm{\hat h}}}_k}} \right)}^2} + 2ab{\bm{\hat h}}_k^{\rm{H}}{{\bm{Q}}_k}^{\frac{1}{2}}{{\bm{R}}^2}{{\bm{Q}}_k}^{\frac{1}{2}}{{{\bm{\hat h}}}_k}{\bm{\bar g}}_{1,k}^{\rm{H}}{{{\bm{\bar g}}}_{1,k}}} \right]\\
\quad +E\left[ { 2ac{\bm{\hat h}}_k^{\rm{H}}{{\bm{Q}}_k}^{\frac{1}{2}}{{\bm{R}}^2}{{\bm{Q}}_k}^{\frac{1}{2}}{{{\bm{\hat h}}}_k}{\bm{\hat h}}_k^{\rm{H}}{{\bm{Q}}_k}^{\frac{1}{2}}{\bm{R}}{{{\bm{\bar g}}}_{1,k}} + 2ac{\bm{\hat h}}_k^{\rm{H}}{{\bm{Q}}_k}^{\frac{1}{2}}{{\bm{R}}^2}{{\bm{Q}}_k}^{\frac{1}{2}}{{{\bm{\hat h}}}_k}{\bm{\bar g}}_{1,k}^{\rm{H}}{\bm{R}}{{\bm{Q}}_k}^{\frac{1}{2}}{{{\bm{\hat h}}}_k}} \right]\\

\quad + E\left[ {{b^2}{{\left( {{\bm{\bar g}}_{1,k}^{\rm{H}}{{{\bm{\bar g}}}_{1,k}}} \right)}^2} + 2bc{\bm{\bar g}}_{1,k}^{\rm{H}}{{{\bm{\bar g}}}_{1,k}}{\bm{\hat h}}_k^{\rm{H}}{{\bm{Q}}_k}^{\frac{1}{2}}{\bm{R}}{{{\bm{\bar g}}}_{1,k}} + 2bc{\bm{\bar g}}_{1,k}^{\rm{H}}{{{\bm{\bar g}}}_{1,k}}{\bm{\bar g}}_{1,k}^{\rm{H}}{\bm{R}}{{\bm{Q}}_k}^{\frac{1}{2}}{{{\bm{\hat h}}}_k}} \right]\\
 \quad + E\left[ {{c^2}{{\left( {{\bm{\hat h}}_k^{\rm{H}}{{\bm{Q}}_k}^{\frac{1}{2}}{\bm{R}}{{{\bm{\bar g}}}_{1,k}}} \right)}^2} + 2{c^2}{\bm{\hat h}}_k^{\rm{H}}{{\bm{Q}}_k}^{\frac{1}{2}}{\bm{R}}{{{\bm{\bar g}}}_{1,k}}{\bm{\bar g}}_{1,k}^{\rm{H}}{\bm{R}}{{\bm{Q}}_k}^{\frac{1}{2}}{{{\bm{\hat h}}}_k}} \right] + E\left[ {{c^2}{{\left( {{\bm{\bar g}}_{1,k}^{\rm{H}}{\bm{R}}{{\bm{Q}}_k}^{\frac{1}{2}}{{{\bm{\hat h}}}_k}} \right)}^2}} \right]
\end{array}\]
where $a \buildrel \Delta \over = {\left( {\frac{{{\lambda _{1,k}}}}{{{\vartheta _k} + 1}}} \right)^2}$, $b \buildrel \Delta \over = \frac{{{\vartheta _k}}}{{{\vartheta _k} + 1}}$, $c \buildrel \Delta \over = \frac{{{\lambda _{1,k}}\sqrt {{\vartheta _k}} }}{{{{\left( {{\vartheta _k} + 1} \right)}^{{3 \mathord{\left/
 {\vphantom {3 2}} \right.
 \kern-\nulldelimiterspace} 2}}}}}$.

 Since ${h_{k,i}},i = 1, \cdots N$ are independent standard complex Gaussian variables, $E\left[ {{\bm{\bar g}}_{1,k}^{\rm{H}}{{{\bm{\bar g}}}_{1,k}}{\bm{\hat h}}_k^{\rm{H}}{{\bm{Q}}_k}^{\frac{1}{2}}{\bm{R}}{{{\bm{\bar g}}}_{1,k}}} \right] = 0$, $E\left[ {{\bm{\bar g}}_{1,k}^{\rm{H}}{{{\bm{\bar g}}}_{1,k}}{\bm{\bar g}}_{1,k}^{\rm{H}}{\bm{R}}{{\bm{Q}}_k}^{\frac{1}{2}}{{{\bm{\hat h}}}_k}} \right] = 0$, and \[E\left[ {{\bm{\hat h}}_k^{\rm{H}}{{\bm{Q}}_k}^{\frac{1}{2}}{\bm{R}}{{{\bm{\bar g}}}_{1,k}}{\bm{\bar g}}_{1,k}^{\rm{H}}{\bm{R}}{{\bm{Q}}_k}^{\frac{1}{2}}{{{\bm{\hat h}}}_k}} \right] = tr\left( {{{\bm{Q}}_k}^{\frac{1}{2}}{\bm{R}}{{{\bm{\bar g}}}_{1,k}}{\bm{\bar g}}_{1,k}^{\rm{H}}{\bm{R}}{{\bm{Q}}_k}^{\frac{1}{2}}} \right){\rm{ = }}{\bm{\bar g}}_{1,k}^{\rm{H}}{\bm{U\Delta }}_k^2{{\bm{U}}^{\rm{H}}}{{\bm{\bar g}}_{1,k}}\]
let ${h_{k,i}} = {s_{k,i}} + j{t_{k,i}}$, then
\[\begin{array}{l}
E\left[ {h_{k,i}^*{{\left| {{h_{k,i}}} \right|}^2}} \right] = E\left[ {\left( {{s_{k,i}} - j{t_{k,i}}} \right)\left( {s_{k,i}^2 + t_{k,i}^2} \right)} \right]\\
\quad \quad \quad \quad \quad \;\; = E\left[ {s_{k,i}^3 + {s_{k,i}}t_{k,i}^2 - j{t_{k,i}}s_{k,i}^2 - jt_{k,i}^3} \right]\\
\quad \quad \quad \quad \quad \;\; = 0
\end{array}\]
Similarly, $E\left[ {h_{k,i}^*{h_{k,i'}}{h_{k,i''}}} \right]\; = 0$, $E\left[ {h_{k,i'}^*{{\left| {{h_{k,i}}} \right|}^2}} \right]\; = 0$, then
\[E\left[ {{\bm{\hat h}}_k^{\rm{H}}{{\bm{Q}}_k}^{\frac{1}{2}}{{\bm{R}}^2}{{\bm{Q}}_k}^{\frac{1}{2}}{{{\bm{\hat h}}}_k}{\bm{\hat h}}_k^{\rm{H}}{{\bm{Q}}_k}^{\frac{1}{2}}{\bm{R}}{{{\bm{\bar g}}}_{1,k}}} \right] = 0\]
\[E\left[ {{\bm{\hat h}}_k^{\rm{H}}{{\bm{Q}}_k}^{\frac{1}{2}}{{\bm{R}}^2}{{\bm{Q}}_k}^{\frac{1}{2}}{{{\bm{\hat h}}}_k}{\bm{\bar g}}_{1,k}^{\rm{H}}{\bm{R}}{{\bm{Q}}_k}^{\frac{1}{2}}{{{\bm{\hat h}}}_k}} \right] = 0\]
and since $E\left[ {{{\left( {h_{k,i}^*} \right)}^2}} \right]\; = E\left[ {s_{k,i}^2 - t_{k,i}^2 - j2{t_{k,i}}{s_{k,i}}} \right]\; = 0$,
\[E\left[ {{{\left( {{\bm{\hat h}}_k^{\rm{H}}{{\bm{Q}}_k}^{\frac{1}{2}}{\bm{R}}{{{\bm{\bar g}}}_{1,k}}} \right)}^2}} \right] = 0\]
\[E\left[ {{{\left( {{\bm{\bar g}}_{1,k}^{\rm{H}}{\bm{R}}{{\bm{Q}}_k}^{\frac{1}{2}}{{{\bm{\hat h}}}_k}} \right)}^2}} \right] = 0\]
Based on the above derivation, the power of the desired signal can be simplified as
\[\begin{array}{l}
\quad E\left[ {{{\left| {{\bm{\hat g}}_{1,k}^{\rm{H}}{{{\bm{\hat g}}}_{1,k}}} \right|}^2}} \right]\\
 = E\left[ {{a^2}{{\left( {{\bm{\hat h}}_k^{\rm{H}}{{\bm{Q}}_k}^{\frac{1}{2}}{{\bm{R}}^2}{{\bm{Q}}_k}^{\frac{1}{2}}{{{\bm{\hat h}}}_k}} \right)}^2} + 2ab{\bm{\hat h}}_k^{\rm{H}}{{\bm{Q}}_k}^{\frac{1}{2}}{{\bm{R}}^2}{{\bm{Q}}_k}^{\frac{1}{2}}{{{\bm{\hat h}}}_k}{\bm{\bar g}}_{1,k}^{\rm{H}}{{{\bm{\bar g}}}_{1,k}} + {b^2}{{\left( {{\bm{\bar g}}_{1,k}^{\rm{H}}{{{\bm{\bar g}}}_{1,k}}} \right)}^2} } \right]\\
 \quad \quad \quad +E\left[ { 2{c^2}{\bm{\bar g}}_{1,k}^{\rm{H}}{\bm{U\Delta }}_k^2{{\bm{U}}^{\rm{H}}}{{{\bm{\bar g}}}_{1,k}}} \right]
\end{array}\]
and \[E\left[ {{{\left( {{\bm{\hat h}}_k^{\rm{H}}{{\bm{Q}}_k}^{\frac{1}{2}}{{\bm{R}}^2}{{\bm{Q}}_k}^{\frac{1}{2}}{{{\bm{\hat h}}}_k}} \right)}^2}} \right] = E\left[ {{{\left( {{\bm{\hat h}}_k^{\rm{H}}{\bm{U\Delta }}_k^2{{\bm{U}}^{\rm{H}}}{{{\bm{\hat h}}}_k}} \right)}^2}} \right]\]
Because ${\bm{U}}$ is unitary matrix, so ${{\bm{\hat h'}}_k} \buildrel \Delta \over = {{\bm{U}}^{\rm{H}}}{{\bm{\hat h}}_k}$Óë${{\bm{\hat h}}_k}$ has the same statistical performances, which means ${{\bm{\hat h'}}_k}$ is standard complex Gaussian vector. So
\[E\left[ {{{\left( {{\bm{\hat h}}_k^{\rm{H}}{{\bm{Q}}_k}^{\frac{1}{2}}{{\bm{R}}^2}{{\bm{Q}}_k}^{\frac{1}{2}}{{{\bm{\hat h}}}_k}} \right)}^2}} \right] = E\left[ {{{\left( {\left({{\bm{\hat h'}}_k}\right)^{\rm{H}}{\bm{\Delta }}_k^2{{{\bm{\hat h'}}}_k}} \right)}^2}} \right] = E\left[ {{{\left( {\sum\limits_{n = 1}^N {\delta_{k,n}^2{{\left| {{{{\bm{\hat h'}}}_{k,n}}} \right|}^2}} } \right)}^2}} \right]\]
According to Lemmas \ref{l3} and \ref{l4}, we have $\delta _{k,n}^2{\left| {{{{\bm{\hat h'}}}_{k,n}}} \right|^2} \sim Gamma\left( {1,\delta _{k,n}^2} \right)$, and based on Lemma \ref{l6}, we have
\[E\left[ {{{\left( {{\bm{\hat h}}_k^{\rm{H}}{{\bm{Q}}_k}^{\frac{1}{2}}{{\bm{R}}^2}{{\bm{Q}}_k}^{\frac{1}{2}}{{{\bm{\hat h}}}_k}} \right)}^2}} \right] = E\left[ {{{\left( {\sum\limits_{n = 1}^N {\delta _{k,n}^2{{\left| {{{{\bm{\hat h'}}}_{k,n}}} \right|}^2}} } \right)}^2}} \right] = \sum\limits_{n = 1}^N {\delta _{k,n}^4}  + {\left( {\sum\limits_{n = 1}^N {\delta _{k,n}^2} } \right)^2}\]
Since ${{\bm{\bar g}}_{1,k}}{\rm{ = }}\sqrt {{\lambda _{1,k}}} {\left[ {1\quad {e^{ - j\frac{{2\pi d}}{\lambda }\sin {\theta _k}}} \cdots {e^{ - j\left( {N - 1} \right)\frac{{2\pi d}}{\lambda }\sin {\theta _k}}}} \right]^{\rm{T}}}$, then ${\bm{\bar g}}_{1,k}^{\rm{H}}{{\bm{\bar g}}_{1,k}}{\rm{ = }}N{\lambda _{1,k}}$£¬
\[E\left[ {{\bm{\hat h}}_k^{\rm{H}}{{\bm{Q}}_k}^{\frac{1}{2}}{{\bm{R}}^2}{{\bm{Q}}_k}^{\frac{1}{2}}{{{\bm{\hat h}}}_k}{\bm{\bar g}}_{1,k}^{\rm{H}}{{{\bm{\bar g}}}_{1,k}}} \right] = E\left[ {\sum\limits_{n = 1}^N {\delta _{k,n}^2{{\left| {{{{\bm{\hat h'}}}_{k,n}}} \right|}^2}} } \right]N{\lambda _{1,k}} = N{\lambda _{1,k}}\sum\limits_{n = 1}^N {\delta _{k,n}^2} \]
\[E\left[ {{{\left( {{\bm{\bar g}}_{1,k}^{\rm{H}}{{{\bm{\bar g}}}_{1,k}}} \right)}^2}} \right] = {N^2}\lambda _{1,k}^2\]

In summary, the power of the desired signal is
\[\begin{array}{l}
 E\left[ {{{\left| {{\bm{\hat g}}_{1,k}^{\rm{H}}{{{\bm{\hat g}}}_{1,k}}} \right|}^2}} \right]
 = {\left( {\frac{{{\lambda _{1,k}}}}{{{\vartheta _k} + 1}}} \right)^4}\left( {\sum\limits_{n = 1}^N {\delta _{k,n}^4}  + {{\left( {\sum\limits_{n = 1}^N {\delta _{k,n}^2} } \right)}^2}} \right) + {\left( {\frac{{{\vartheta _k}}}{{{\vartheta _k} + 1}}} \right)^2}{N^2}\lambda _{1,k}^2\\
\quad \quad \quad \quad \quad \quad \quad \quad + 2\frac{{\lambda _{1,k}^2{\vartheta _k}}}{{{{\left( {{\vartheta _k} + 1} \right)}^3}}}\left[ {N{\lambda _{1,k}}\sum\limits_{n = 1}^N {\delta _{k,n}^2}  + {\bm{\bar g}}_{1,k}^{\rm{H}}{\bm{U\Delta }}_k^2{{\bm{U}}^{\rm{H}}}{{{\bm{\bar g}}}_{1,k}}} \right]
\end{array}\]

2. The power of interfering signal and noise
\begin{align}\label{eq:sigma}
&\quad E\left[ {{\bm{\hat g}}_{1,k}^{\rm{H}}\left( {\sum\limits_{l = 1}^L {\sum\limits_{i \ne k}^K {{{{\bm{\hat g}}}_{l,i}}} {\bm{\hat g}}_{l,i}^{\rm{H}}}  + \sum\limits_{l = 1}^L {\sum\limits_{i = 1}^K {{{{\bm{\tilde g}}}_{l,i}}{\bm{\tilde g}}_{l,i}^{\rm{H}}} }  + \frac{{{{\bm{I}}_N}}}{{{p_u}}} + \sum\limits_{l = 2}^L {{{{\bm{\hat g}}}_{l,k}}{\bm{\hat g}}_{l,k}^{\rm{H}}} } \right){{{\bm{\hat g}}}_{1,k}}} \right] \nonumber \\
 &= E\left[ {E\left[ {{\bm{\hat g}}_{1,k}^{\rm{H}}\left( {\sum\limits_{l = 1}^L {\sum\limits_{i \ne k}^K {{{{\bm{\hat g}}}_{l,i}}} {\bm{\hat g}}_{l,i}^{\rm{H}}}  + \sum\limits_{l = 1}^L {\sum\limits_{i = 1}^K {{{{\bm{\tilde g}}}_{l,i}}{\bm{\tilde g}}_{l,i}^{\rm{H}}} }  + \frac{{{{\bm{I}}_N}}}{{{p_u}}} + \sum\limits_{l = 2}^L {{{{\bm{\hat g}}}_{l,k}}{\bm{\hat g}}_{l,k}^{\rm{H}}} } \right){{{\bm{\hat g}}}_{1,k}}\left| {{{{\bm{\hat g}}}_{l,k}}\left( {l = 1, \cdots ,L} \right)} \right.} \right]} \right] \nonumber \\
& = E\left[ {{\bm{\hat g}}_{1,k}^{\rm{H}}\left( {{\bm{\Sigma }} + \frac{{{{\bm{I}}_N}}}{{{p_u}}} + \sum\limits_{l = 2}^L {{{{\bm{\hat g}}}_{l,k}}{\bm{\hat g}}_{l,k}^{\rm{H}}} } \right){{{\bm{\hat g}}}_{1,k}}} \right]
\end{align}
where
\begin{align}
{\bm{\Sigma }} &= \sum\limits_{i = 1}^K {\left( {\frac{{{\lambda _{1,i}}{\bm{R}}}}{{{\vartheta _i} + 1}}} \right)}  + \sum\limits_{l = 2}^L {\sum\limits_{i = 1}^K {{\lambda _{l,i}}{\bm{R}}} }  - {\left( {\frac{{{\lambda _{1,k}}}}{{{\vartheta _k} + 1}}} \right)^2}{\bm{R}}{{\bm{Q}}_k}{\bm{R}} - \sum\limits_{l = 2}^L {\lambda _{l,k}^2{\bm{R}}{{\bm{Q}}_k}{\bm{R}}}  + \sum\limits_{i \ne k}^K {\left[ {\frac{{{\vartheta _i}}}{{{\vartheta _i} + 1}}{{{\bm{\bar g}}}_{1,i}}{\bm{\bar g}}_{1,i}^{\rm{H}}} \right]} \\ \nonumber
& \buildrel \Delta \over={\bm{U{\rm A}}}{{\bm{U}}^{\rm{H}}} + \sum\limits_{i \ne k}^K {\left[ {\frac{{{\vartheta _i}}}{{{\vartheta _i} + 1}}{{{\bm{\bar g}}}_{1,i}}{\bm{\bar g}}_{1,i}^{\rm{H}}} \right]}
 \end{align}
where\[{\bm{A}} \buildrel \Delta \over = \sum\limits_{i = 1}^K {\left( {\frac{{{\lambda _{1,i}}{\bm{D}}}}{{{\vartheta _i} + 1}}} \right)}  + \sum\limits_{l = 2}^L {\sum\limits_{i = 1}^K {{\lambda _{l,i}}{\bm{D}}} }  - {\left( {\frac{{{\lambda _{1,k}}}}{{{\vartheta _k} + 1}}} \right)^2}{{\bm{\Delta }}_k^2} - \sum\limits_{l = 2}^L {\lambda _{l,k}^2{{\bm{\Delta }}_k^2}} \]

The expectation of \eqref{eq:sigma} is divided into three parts£º

(a) \[\begin{array}{l}
E\left[ {{\bm{\hat g}}_{1,k}^{\rm{H}}{\bm{\Sigma }}{{{\bm{\hat g}}}_{1,k}}} \right]= {\left( {\frac{{{\lambda _{1,k}}}}{{{\vartheta _k} + 1}}} \right)^2}tr\left( {{\bm{{\rm A}\Delta }}_k^2} \right) + {\left( {\frac{{{\lambda _{1,k}}}}{{{\vartheta _k} + 1}}} \right)^2}\sum\limits_{i \ne k}^K {\left[ {\frac{{{\vartheta _i}}}{{{\vartheta _i} + 1}}{\bm{\bar g}}_{1,i}^{\rm{H}}{\bm{R}}{{\bm{Q}}_k}{\bm{R}}{{{\bm{\bar g}}}_{1,i}}} \right]} \\
 \quad \quad \quad  \quad \quad \quad  \quad \quad + \frac{{{\vartheta _k}}}{{{\vartheta _k} + 1}}{\bm{\bar g}}_{1,k}^{\rm{H}}\left( {{\bm{U{\rm A}}}{{\bm{U}}^{\rm{H}}} + \sum\limits_{i \ne k}^K {\left[ {\frac{{{\vartheta _i}}}{{{\vartheta _i} + 1}}{{{\bm{\bar g}}}_{1,i}}{\bm{\bar g}}_{1,i}^{\rm{H}}} \right]} } \right){{{\bm{\bar g}}}_{1,k}}
\end{array}\]
(b) Similar to the proof of the power of desired signal,
\[\begin{array}{l}
\quad E\left[ {{\bm{\hat g}}_{1,k}^{\rm{H}}\left( {\sum\limits_{l = 2}^L {{{{\bm{\hat g}}}_{l,k}}{\bm{\hat g}}_{l,k}^{\rm{H}}} } \right){{{\bm{\hat g}}}_{1,k}}} \right]\\
 = \sum\limits_{l = 2}^L {\frac{{\lambda _{1,k}^2\lambda _{l,k}^2}}{{{{\left( {{\vartheta _k} + 1} \right)}^2}}}\left( {\sum\limits_{n = 1}^N {\delta _{k,n}^4}  + {{\left( {\sum\limits_{n = 1}^N {\delta _{k,n}^2} } \right)}^2}} \right)}  + \frac{{{\vartheta _k}}}{{{\vartheta _k} + 1}}\left( {{\bm{\bar g}}_{1,k}^{\rm{H}}{\bm{U\Delta }}_k^2{{\bm{U}}^{\rm{H}}}{{{\bm{\bar g}}}_{1,k}}} \right)\sum\limits_{l = 2}^L {\lambda _{l,k}^2}
\end{array}\]
(c) The power of the noise
\[E\left[ {\frac{{{\bm{\hat g}}_{1,k}^{\rm{H}}{{{\bm{\hat g}}}_{1,k}}}}{{{p_u}}}} \right] = \frac{1}{{{p_u}}}\left[ {{{\left( {\frac{{{\lambda _{1,k}}}}{{{\vartheta _k} + 1}}} \right)}^2}\sum\limits_{n = 1}^N {\delta _{k,n}^2} + \frac{{{\vartheta _k}{\lambda _{1,k}}N}}{{{\vartheta _k} + 1}}} \right]\]

In order to facilitate the analysis of the impact of LOS component and the Rayleigh component to the uplink rate, we divide the total power into two parts: the LOS related power and the Rayleigh related power.
\end {IEEEproof}

\subsection{Proof of Theorem \ref{thm:lmmsei}}\label{mcell_dl_1}
\begin{IEEEproof}
First we introduce extreme-value questions of four functions that we will use in the following proof:
If $\sum\limits_{m = 1}^M {{x_m}}  = M$, for any $a > 0,\;b > 0$, ${x_m} > 0$,  we have
\begin{equation}\label{eq:inequa1}\frac{{{M^2}}}{{aM + b}} > \sum\limits_{m = 1}^M {\frac{{x_m^2}}{{a{x_m} + b}}}  \ge \frac{M}{{a + b}}\end{equation}
\begin{equation}\label{eq:inequa2}\frac{{{M^3}}}{{aM + b}} > \sum\limits_{m = 1}^M {\frac{{x_m^3}}{{a{x_m} + b}}}  \ge \frac{M}{{a + b}}\end{equation}
\begin{equation}\label{eq:inequa3}\frac{{{M^4}}}{{{{\left( {aM + b} \right)}^2}}} > \sum\limits_{m = 1}^M {\frac{{x_m^4}}{{{{\left( {a{x_m} + b} \right)}^2}}}}  \ge \frac{M}{{{{\left( {a + b} \right)}^2}}}\end{equation}
\begin{equation}\label{eq:inequa4}\frac{{{M^3}}}{{{{\left( {aM + b} \right)}^2}}} > \sum\limits_{m = 1}^M {\frac{{x_m^3}}{{{{\left( {a{x_m} + b} \right)}^2}}}}  \ge \frac{M}{{{{\left( {a + b} \right)}^2}}}\end{equation}
and the upper limits of the above four inequality are achieved when ${x_i}=M$ and ${x_m}=0,m\ne i$, while the lower limits are achieved when ${x_i}=1,\forall i $.

We separate the proof of asymptotic expression of $k$th user's SINR into three parts:

1.
\[\delta _{k,n}^2 = \frac{{d_n^2}}{{\left( {\frac{{{\lambda _{1,k}}}}{{{\vartheta _k} + 1}} + \sum\limits_{l = 2}^L {{\lambda _{l,k}}} } \right){d_n} + \frac{1}{{{p_P}}}}}\]
since $\bm{R}$ has the following properties:
positive definite; $\text{Tr}\left[ \bm{R} \right] = N$; has uniformly bounded spectral norm. So $ \frac{{\sum\limits_{n = 1}^N {\delta _{k,n}^2} }}{N} $ and $\frac{{\sum\limits_{n = 1}^N {\delta _{k,n}^4} }}{{{N}}}$ are non-zero finite values. When $N \to \infty $, $\frac{{\sum\limits_{n = 1}^N {\delta _{k,n}^4} }}{{{N^2}}}\to 0$.

And after some algebraic manipulations, we have \[{a_n}\delta _{k,n}^2 = \frac{{{\tau _k}d_n^4}}{{{{\left[ {\left( {\frac{{{\lambda _{1,k}}}}{{{\vartheta _k} + 1}} + \sum\limits_{l = 2}^L {{\lambda _{l,k}}} } \right){d_n} + \frac{1}{{{p_P}}}} \right]}^2}}} + \frac{{\frac{{\sum\limits_{i = 1}^K {\left( {\frac{{{\lambda _{1,i}}}}{{{\vartheta _i} + 1}}} \right)}  + \sum\limits_{l = 2}^L {\sum\limits_{i = 1}^K {{\lambda _{l,i}}} } }}{{{p_P}}}d_n^3}}{{{{\left[ {\left( {\frac{{{\lambda _{1,k}}}}{{{\vartheta _k} + 1}} + \sum\limits_{l = 2}^L {{\lambda _{l,k}}} } \right){d_n} + \frac{1}{{{p_P}}}} \right]}^2}}}\]
where \[{\tau _k} \buildrel \Delta \over = \left[ {\sum\limits_{i \ne k}^K {\left( {\frac{{{\lambda _{1,i}}}}{{{\vartheta _i} + 1}}} \right)}  + \sum\limits_{l = 2}^L {\sum\limits_{i \ne k}^K {{\lambda _{l,i}}} } } \right]\left( {\frac{{{\lambda _{1,k}}}}{{{\vartheta _k} + 1}} + \sum\limits_{l = 2}^L {{\lambda _{l,k}}} } \right) + \frac{{{\rm{2}}{\lambda _{1,k}}\sum\limits_{l = 2}^L {{\lambda _{l,k}}} }}{{{\vartheta _k} + 1}} > 0\]
Similarly, when $N \to \infty $, $\frac{{\sum\limits_{n = 1}^N {{a_n}\delta _{k,n}^2} }}{N^2} \to 0$.

2.
\[{\bm{\bar g}}_{1,k}^{\rm{H}}{{\bm{\bar g}}_{1,i}}{\bm{\bar g}}_{1,i}^{\rm{H}}{{\bm{\bar g}}_{1,k}} = \left\{ \begin{array}{ll}
\lambda _{1,k}^2{N^2}  & k = i\\
\lambda _{1,k}^{}\lambda _{1,i}{{{\left| {{\rho _{k,i}}} \right|}^2}}& k \ne i
\end{array} \right.\]
where ${\rho _{k,i}} = \frac{{1 - {e^{jN{\varphi _{ki}}}}}}{{1 - {e^{j{\varphi _{ki}}}}}}$, ${\varphi _{ki}} = \frac{{2\pi d}}{\lambda }\left( {\sin {\theta _k} - \sin {\theta _i}} \right)$.
Because ${\left| {1 - {e^{jN{\varphi _{ki}}}}} \right|}$ is finite, so when $N \to \infty $, \[\frac{{{\bm{\bar g}}_{1,k}^{\rm{H}}{{{\bm{\bar g}}}_{1,i}}{\bm{\bar g}}_{1,i}^{\rm{H}}{{{\bm{\bar g}}}_{1,k}}}}{{{N^2}}} \to \left\{ \begin{array}{l}
\lambda _{1,k}^2\quad k = i\\
\;\,{\rm{0}}\;\quad \;k \ne i
\end{array} \right.\]

3.
Define ${\bm{\bar g}}_{1,k}^{\rm{H}}{\bm{U}}{\rm{ = }}{\bm{v}} = \left( {{v_1}\;{v_2} \cdots {v_N}} \right)$, then ${\bm{\bar g}}_{1,k}^{\rm{H}}{\bm{U\Delta }}_k^2{{\bm{U}}^{\rm{H}}}{{\bm{\bar g}}_{1,k}} = \sum\limits_{n = 1}^N {\delta _{k,n}^2{{\left| {{v_n}} \right|}^2}} $, and let ${\delta _{\min }} = \mathop {\min }\limits_{n = 1, \cdots ,N} \delta _{k,n}^2$, ${\delta _{\max }} = \mathop {\max }\limits_{n = 1, \cdots ,N} \delta _{k,n}^2$, then \[{\delta _{\min }}\sum\limits_{n = 1}^N {{{\left| {{v_n}} \right|}^2}}  \le {\bm{\bar g}}_{1,k}^{\rm{H}}{\bm{U\Delta }}_k^2{{\bm{U}}^{\rm{H}}}{{\bm{\bar g}}_{1,k}} \le {\delta _{\max }}\sum\limits_{n = 1}^N {{{\left| {{v_n}} \right|}^2}} \]

Since $\sum\limits_{n = 1}^N {{{\left| {{v_n}} \right|}^2}} {\rm{ = }}{\bm{\bar g}}_{1,k}^{\rm{H}}{\bm{U}}{{\bm{U}}^{\rm{H}}}{{\bm{\bar g}}_{1,k}}{\rm{ = }}N{\lambda _{1,k}}$, \[{\delta _{\min }}N{\lambda _{1,k}} \le {\bm{\bar g}}_{1,k}^{\rm{H}}{\bm{U\Delta }}_k^2{{\bm{U}}^{\rm{H}}}{{\bm{\bar g}}_{1,k}} \le {\delta _{\max }}N{\lambda _{1,k}}\]
Because $\bm{R}$ has uniformly bounded spectral norm, so ${\delta _{\min }}$, ${\delta _{\max }}$ is finite when $N \to \infty $, \[\frac{{{\bm{\bar g}}_{1,k}^{\rm{H}}{\bm{U\Delta }}_k^2{{\bm{U}}^{\rm{H}}}{{{\bm{\bar g}}}_{1,k}}}}{{{N^{\rm{2}}}}} \to {\rm{0}}\]

And since \[{b_n} = \left[ {\sum\limits_{i = 1}^K {\left( {\frac{{{\lambda _{l,i}}}}{{{\vartheta _i} + 1}}} \right)}  + \sum\limits_{l = 2}^L {\sum\limits_{i = 1}^K {{\lambda _{l,i}}} } } \right]{d_n} - \frac{{\lambda _{1,k}^2}}{{{{\left( {{\vartheta _k} + 1} \right)}^2}}}\frac{{d_n^2}}{{\frac{{{\lambda _{1,k}}{d_n}}}{{{\vartheta _k} + 1}} + \sum\limits_{l = 2}^L {{\lambda _{l,k}}{d_n}}  + \frac{1}{{{p_P}}}}}\]
similarly, we have \[\frac{{{\bm{\bar g}}_{1,k}^{\rm{H}}{\bm{UB}}{{\bm{U}}^{\rm{H}}}{{{\bm{\bar g}}}_{1,k}}}}{{{N^{\rm{2}}}}} \to {\rm{0}}\]

Finally, based on the above three conclusions, when $N\to \infty$, we have
\[\frac{{{S_{LOS,k}}}}{{{N^2}}} \to \frac{{\lambda _{1,k}^2}}{{{{\left( {{\vartheta _k} + 1} \right)}^2}}}\left[ {2\frac{{{\vartheta _k}{\lambda _{1,k}}}}{{{\vartheta _k} + 1}}\left( {\frac{{\sum\limits_{n = 1}^N {\delta _{k,n}^2} }}{N}} \right) + \vartheta _k^2} \right]\]
\[\frac{{{S_{w,k}}}}{{{N^2}}} \to \frac{{\lambda _{1,k}^2}}{{{{\left( {{\vartheta _k} + 1} \right)}^2}}}{\left( {\frac{{{\lambda _{1,k}}}}{{{\vartheta _k} + 1}}} \right)^2}{\left( {\frac{{\sum\limits_{n = 1}^N {\delta _{k,n}^2} }}{N}} \right)^2}\]
\[\frac{{{I_{w,k}}}}{{{N^2}}} \to \frac{{\lambda _{1,k}^2}}{{{{\left( {{\vartheta _k} + 1} \right)}^2}}}\left[ {\sum\limits_{l = 2}^L {\lambda _{l,k}^2{{\left( {\frac{{\sum\limits_{n = 1}^N {\delta _{k,n}^2} }}{N}} \right)}^2}} } \right]\]
\[\frac{{{I_{LOS,k}}}}{{{N^2}}} \to 0\]
and after some algebraic operations
\[SIN{R_k} \to \frac{{{{\left( {\frac{{{\lambda _{1,k}}}}{{{\vartheta _k} + 1}} + \frac{{{\vartheta _k}}}{{{{\left( {\sum\limits_{n = 1}^N {\delta _{k,n}^2} } \right)} \mathord{\left/
 {\vphantom {{\left( {\sum\limits_{n = 1}^N {\delta _{k,n}^2} } \right)} N}} \right.
 \kern-\nulldelimiterspace} N}}}} \right)}^2}}}{{\sum\limits_{l = 2}^L {\lambda _{l,k}^2} }}\buildrel \Delta \over ={{\widehat  {SINR} }_k^{\infty}} \]
\end{IEEEproof}

\subsection{Proof of Theorem \ref{powersl}}\label{z4}
\begin{IEEEproof}
Substituting ${p_u} = {E_u}{N^{ - \varepsilon }}$ into
\[\delta _{k,n}^2 = \frac{{d_n^2}}{{\left( {\frac{{{\lambda _{1,k}}}}{{{\vartheta _k} + 1}} + \sum\limits_{l = 2}^L {{\lambda _{l,k}}} } \right){d_n} + \frac{1}{{K{p_u}}}}}\]
and when $N \to \infty $, \[\delta _{k,n}^2 \to K{E_u}{N^{ - \varepsilon }}d_n^2\]
\[{\bm{\Delta }}_k^2 \to K{E_u}{N^{ - \varepsilon }}{{\bm{D}}^2}\]
\[\sum\limits_{n = 1}^N {\delta _{k,n}^2}  \to K{E_u}{N^{ - \varepsilon }}\sum\limits_{n = 1}^N {d_n^2} \]
\[\sum\limits_{n = 1}^N {\delta _{k,n}^{\rm{4}}}  \to K{E_u}{N^{ - \varepsilon }}\sum\limits_{n = 1}^N {d_n^{\rm{4}}} \]
\[{a_n}\delta _{k,n}^2 \to K{E_u}{N^{ - \varepsilon }}\left[ {\sum\limits_{i = 1}^K {\left( {\frac{{{\lambda _{1,i}}}}{{{\vartheta _i} + 1}}} \right)}  + \sum\limits_{l = 2}^L {\sum\limits_{i = 1}^K {{\lambda _{l,i}}} } } \right]d_n^3 - \left[ {\frac{{\lambda _{1,k}^2}}{{{{\left( {{\vartheta _k} + 1} \right)}^2}}} + \sum\limits_{l = 2}^L {\lambda _{l,k}^2} } \right]{\left( {K{E_u}{N^{ - \varepsilon }}} \right)^2}d_n^4\]
\[{b_n} \to \left[ {\sum\limits_{i = 1}^K {\left( {\frac{{{\lambda _{1,i}}}}{{{\vartheta _i} + 1}}} \right)}  + \sum\limits_{l = 2}^L {\sum\limits_{i = 1}^K {{\lambda _{l,i}}} } } \right]{d_n} - \frac{{\lambda _{1,k}^2}}{{{{\left( {{\vartheta _k} + 1} \right)}^2}}}K{E_u}{N^{ - \varepsilon }}d_n^2\]
According to the following properties of $\bm{R}$: positive definite; $\text{Tr}\left[ \bm{R} \right] = N$; has uniformly bounded spectral norm,
$\frac{{\sum\limits_{n = 1}^N {d_n^2} }}{N},\frac{{\sum\limits_{n = 1}^N {d_n^3} }}{N},\frac{{\sum\limits_{n = 1}^N {d_n^4} }}{N}$ have non-zero finite values. So \[\frac{{{{\left( {{\vartheta _k} + 1} \right)}^2}}}{{{N^2}\lambda _{1,k}^2}}{S_{LOS,k}} \to \vartheta _k^2\]
\[\frac{{{{\left( {{\vartheta _k} + 1} \right)}^2}}}{{{N^2}\lambda _{1,k}^2}}{S_{w,k}} \to 0\]
\[\frac{{{{\left( {{\vartheta _k} + 1} \right)}^2}}}{{{N^2}\lambda _{1,k}^2}}{I_{w,k}} \to 0\]
\[\frac{{{{\left( {{\vartheta _k} + 1} \right)}^2}}}{{{N^2}\lambda _{1,k}^2}}{I_{LOS,k}} \to \frac{1}{{{N^{1 - \alpha }}{E_u}{\lambda _{1,k}}}}{\vartheta _k}\left( {{\vartheta _k} + 1} \right)\]
then
\[\overline {SINR} _k^{ps} = \frac{{{N^{1 - \varepsilon }}{E_u}{\lambda _{1,k}}{\vartheta _k}}}{{\left( {{\vartheta _k} + 1} \right)}}\]
If ${\vartheta _k} = 0$, then $S_{LOS,k}=0$, $S_{LOS,k}=0$,
\[\frac{{{S_{w,k}}{N^{2\varepsilon }}}}{{{N^2}}} \to \lambda _{1,k}^4{\left( {\frac{{K{E_u}\sum\limits_{n = 1}^N {d_n^2} }}{N}} \right)^2}\]
\[\frac{{{I_{w,k}}{N^{2\varepsilon }}}}{{{N^2}}} \to \lambda _{1,k}^2\left[ {\sum\limits_{l = 2}^L {\lambda _{l,k}^2{{\left( {\frac{{K{E_u}\sum\limits_{n = 1}^N {d_n^2} }}{N}} \right)}^2} + \frac{{K{N^{2\varepsilon  - 1}}\sum\limits_{n = 1}^N {d_n^2} }}{N}} } \right]\]
´Ó¶ø\[SINR_k^{ps} = \frac{{\lambda _{1,k}^2}}{{\sum\limits_{l = 2}^L {\lambda _{l,k}^2}  + \frac{1}{{KE_u^2{N^{ - 2\varepsilon }}\sum\limits_{n = 1}^N {d_n^2} }}}}\]
\end{IEEEproof}

\section*{Acknowledgment}
This work was supported in part by the National Basic Research Program of China (973 Program 2013CB336600), the Natural Science Foundation of China (NSFC) under grants 61271205, 61501113, 61521061, 61371113, 61221002, 61401241, 61501264, China High-Tech 863 Program under
Grant No.2014AA01A704, No.2014AA01A706 and the Natural Science Foundation of Jiangsu Province under grant BK20150630, the Open Research Fund of National Mobile Communications Research Laboratory, Southeast University under Grant No. 2015D02.

\ifCLASSOPTIONcaptionsoff
  \newpage
\fi

\end{document}